\documentclass[prd,preprint,nofootinbib]{revtex4}
\usepackage{graphicx}
\usepackage{amssymb}

\begin{document}

\title{Constraining Neutrino Masses by CMB Experiments Alone}

\author{Kazuhide Ichikawa, Masataka Fukugita and Masahiro
Kawasaki}\affiliation{Institute for Cosmic Ray Research,
University of Tokyo, Kashiwa 277 8582,
Japan}

\date{\today}

\vskip10mm
\begin{abstract}
It is shown that a subelectronvolt upper limit can be derived
on the neutrino mass from the CMB data alone in the $\Lambda$CDM
model with the power-law adiabatic perturbations, without the aid
of any other cosmological data.
Assuming the flatness of the universe, the constraint we can derive
   from the current WMAP observations is $\sum m_{\nu} <  2.0$ eV
at the 95\% confidence level for the sum over three species of neutrinos
($m_\nu<0.66$ eV for the degenerate neutrinos) by maximising the likelihood over 6 other cosmological parameters.
This constraint modifies little even if we abandon the flatness
assumption for the spatial curvature.
We argue that it would be difficult to improve the limit much beyond
$\sum m_{\nu}\lesssim  1.5$ eV using only the CMB data, even if
their statistics are substantially improved. However, a significant 
improvement
of the limit is possible if an external input is introduced that 
constrains
the Hubble constant from below.
The parameter correlation and the
mechanism of CMB perturbations that give rise to the
limit on the neutrino mass are also elucidated.
\end{abstract}
\maketitle

\section{Introduction} \label{sec:introduction}

The upper limit on the absolute mass of neutrinos is derived from
the end-point spectrum of tritium beta decay experiments. It is
not easy, however, to push the limit to the subelectronvolt range.
An alternative hope is to resort to cosmological considerations.
The presence of massive neutrinos
affects cosmic perturbations, most characteristically in a way to reduce
the power in the small scale due to free streaming in the early
universe. In a low matter density universe the effect is significant
even if the neutrino mass is of the order of subelectronvolts 
\cite{Hu:1997mj},
and constraints of a few eV as upper limits on the sum mass of
three species of neutrinos are obtained from the power of
galaxy clustering combined with the normalisation of the fluctuation
power at large scales from
the magnitude of quadrupole anisotropies in the cosmic microwave 
background
(CMB) temperature field \cite{Croft:1999,Fukugita:2000},
or from the shape of the power spectrum of galaxy clustering
\cite{Elgaroey:2002}.

Massive neutrinos also affect perturbations in the CMB temperature
field at intermediate to small scales in a less trivial manner
(see \cite{Ma:1995ey,Dodelson:1995es} for the earlier work).
The effect here is via the modification of CMB perturbations,
especially through the integrated Sachs-Wolfe effect, rather 
than simply the reduction of the power at small scales.
Combining the CMB multipoles of WMAP
with the galaxy clustering data of 2dFGRS,
Spergel et al.~\cite{Spergel:2003cb}
derived $\sum m_\nu<0.7$ eV: using the SDSS power spectrum,
Tegmark et al. \cite{Tegmark:2003ud} give $<1.7$ eV for the sum mass;
see also  Refs.~\cite{Elgaroy:2003yh,Hannestad:2003xv,Allen:2003pt,
Crotty:2004gm,Seljak:2004xh}.
A general problem with the cosmological
analyses is how the result depends on explicit or implicit
assumptions and systematics, especially when two or more pieces
of different types of data, such as CMB multipoles and galaxy clustering
data, are combined. In this context it is an important question to ask 
whether one can derive a comparable limits
on the neutrino mass from the CMB data alone.
Tegmark et al.'s
analysis shows that such a limit is not derived from the
CMB data (WMAP data) alone, allowing for the possibility
that massive neutrinos represent the entire dark matter at one sigma
confidence level, whereas earlier Eisenstein et al.'s
work \cite{Eisenstein:1998hr} seems to
forecast the contrary. We consider that this is an important
point that deserves further studies, especially in the view
that the quality of the CMB temperature field
data will be improved in the future,
notably by the PLANCK in a half decade time,
and it is a consequential question
if one can improve the limit on the neutrino mass without
resorting to the large-scale galaxy clustering data,
for which we always have a suspect for unknown biasing
and not well-controlled nonlinear effects.

It is also important to understand whether the limit depends
upon the assumption of the exactly flat spatial curvature of the 
universe,
as customarily assumed when the constraints on neutrinos were discussed.
We already know that the curvature is quite close to flat, but
the possibility of a slight departure from the flatness
is not excluded.  For instance, the derivation of the consistent
Hubble constant from CMB alone depends crucially on the flatness
assumption: a slight departure, say by 2\% in the spatial curvature,
largely modifies the ``CMB best value'' of the
Hubble constant to an unacceptably small value. 
We see some reason that a small neutrino mass may give
an effect similar to non-flat curvature and thus the two effects
might cancel, loosening the limit.

In this paper we investigate the problem  within the $\Lambda$CDM 
universe with adiabatic perturbations
whether a sensible
limit on the neutrino mass can be derived from the CMB data alone,
and if this is the case how does the limit depend upon the assumption
of the exact flatness of the universe.
A particular emphasis is given to elucidating the parameter correlation
and the mechanism
in the CMB perturbation theory as to how the neutrino mass
limit is derived. In our argument we extensively use the ``reduced
CMB observables'', the position of the first acoustic peak
$\ell_1$, the height of the first peak normalised to the low
$\ell$ value $H_1$, the height of the second relative
to the first peak $H_2$, and
the height of the third relative to the first peak $H_3$, introduced in
Hu et al. \cite{Hu:2000ti}, and study how the massive neutrinos
affect these variables.

We assume that the three neutrinos have a degenerate mass. This will be 
a
realistic assumption if the neutrinos have masses close to the upper
limit that concerns us, because the neutrino oscillation experiments
tell us that the differences of masses are much smaller than the
upper limit.
In our numerical work we ignore the tensor perturbations, but we argue 
that
their inclusion would only tighten the limit on the neutrino mass.
We assume that the
cosmic density perturbations have a power spectrum specified by index 
$n_s$.
A small departure from the power spectrum as predicted by slow-roll
inflation does not change our analysis. If the departure
is at a large amount, such as that indicated by the WMAP team
combining their CMB data with the galaxy clustering,
our result will need modification:
in such a case one cannot argue for the limit on the neutrino mass 
unless the
primordial power spectrum is given.

In the next section, we show with the numerical work that we can derive
a sensible limit on the neutrino mass from the CMB data alone
under the assumption of the
exact spatial flatness of the universe.
In Sec.~\ref{sec:reason} we consider the effect of massive neutrinos
on the reduced CMB observables, and discuss how
one can obtain the constraint from the CMB data alone.
In Sec.~\ref{sec:analytic} we discuss the physics of the response of
the reduced CMB observables to massive neutrinos in CMB
perturbation theory.
In Sec.~\ref{sec:nonflat}, we consider the constraint in non-flat
universes, and show that a comparable constraint is derived.
The conclusion is given in Sec.~\ref{sec:conclusion}.

\section{Limit on the neutrino mass from WMAP alone} 
\label{sec:analysis}
\begin{figure}
\includegraphics{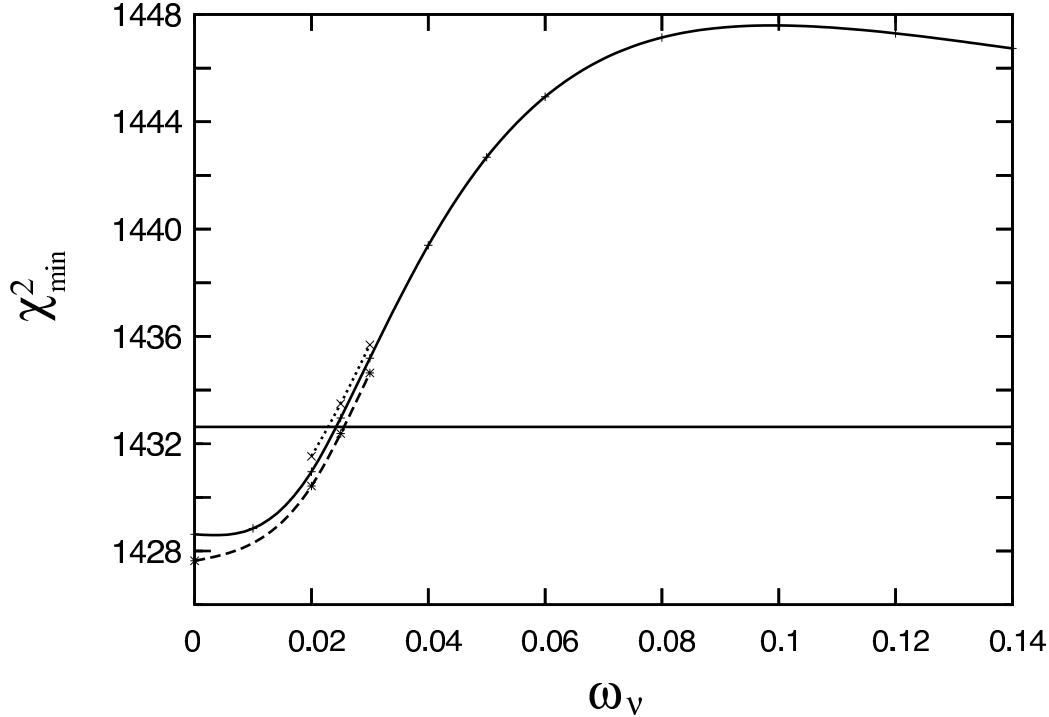}
\caption{Minimum $\chi^2$ as a function of the neutrino energy density
$\omega_{\nu}$. 
The solid curve is for the flat universe. The dotted
and the dashed curves show the cases for a negative and a positive
curvature universe, respectively.}
\label{fig:chi2min}
\end{figure}

\begin{table}
\begin{tabular}{|c|c|c|c|c |c|c||c||c|c |c|c|}
\hline
$\omega_{\nu}$ & $\omega_b$ & $\omega_m$ &   $h$ & $\tau$ & $n_s$ & $A$
& $\chi^2$ &  $\ell_1$ &  $H_1$ &  $H_2$ & $H_3$ \\
\hline
0.00 &0.0230	& 0.145 &			0.689 &	0.116 &	0.973 &	1133.1 &1428.6 &	
	220 &	6.68 &	0.449 &		0.456 \\	
0.001 &0.0231 &	0.145 &			0.682 &	0.116 &	0.973 &	1119.1 &	1428.7
&		220 &		6.70 &		0.449 &		0.459 \\	
0.01 &0.0224 &	0.145 &	0.600 &	0.105 &	0.950 &	1044.7 &		1428.8 &	219
&	6.60 &		0.447 &		0.452 \\	
0.02 &0.0218 &	0.137 &		0.564 &	0.0936 &	0.918 &	1096.7 &	1431.0
&		219	 & 	6.33 &	0.442 &		0.432 \\	
0.025 &0.0216 &	0.130 &		0.556 &	0.0901 &	0.904 &	1137.0	& 1433.0	&
219 &6.21 &	 0.441	& 	0.422	\\
0.03 &0.0219	& 0.128 &		0.551 &	0.0835 &	0.894	&  1155.7 &1435.2
&		219	& 	6.12 &	 0.439 &	 0.417	\\
0.04 &0.0223 &	0.120	& 	0.545 &	0.0793 &	0.883	& 1188.9 & 1439.4	&
	220 &	 6.02 &		0.438 &	 0.411	\\
0.05 &0.0230 &	0.112	& 	0.545 &	0.0794	& 0.876 &	1215.4 &1442.7	&
	220	&  5.95 &	 0.437	& 0.408	\\
0.06 & 0.0237 &	0.104	& 	0.545	& 0.0788	& 0.873 &		1228.1 &	1444.9
&	220 &	 5.92 &		0.437 &		0.406	\\
0.08	&0.0250	& 0.0864	& 	 0.547	& 0.0694	& 0.867 &	 1230.5 &	
1447.1	&	220 &		5.90	&  0.437 &0.402	\\
0.10	&0.0260	& 0.0686	& 	 0.547	& 0.0696 &	0.868	&  1226.2 & 1447.6
&		221 &	 5.90 &	 0.440 &		0.401	\\
0.12 &0.0270	& 0.0496	& 		0.548 &	0.0720 &	0.869 &	1226.2 & 1447.3	&
	221	&  5.92 &	0.441 &	 0.399	\\
0.14 & 0.0278 &	0.0310	& 	0.548 &	0.0741	& 0.874 &		 1213.0
&1446.7	&	221	& 5.95 &	 0.443 &	 0.401	\\
\hline
\end{tabular}
\caption{Solutions for $\chi^2_{\rm min}(\omega_{\nu})$.}
\label{tab:chi2min}
\end{table}

\begin{figure}
\begin{center}
\begin{tabular}{cc}
\includegraphics[width=7.5cm] {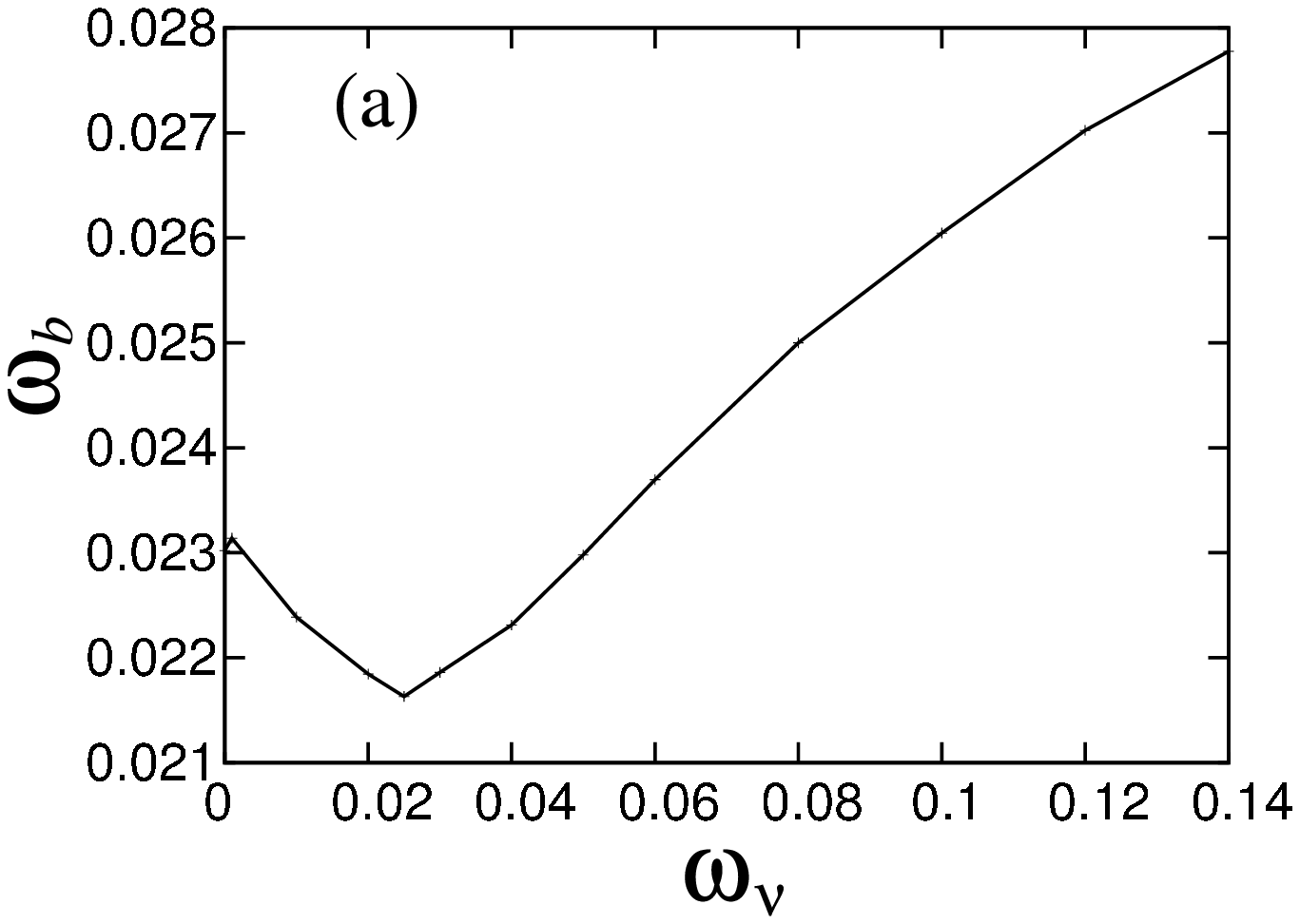} &
\hspace{0.5cm}\includegraphics[width=7.5cm] {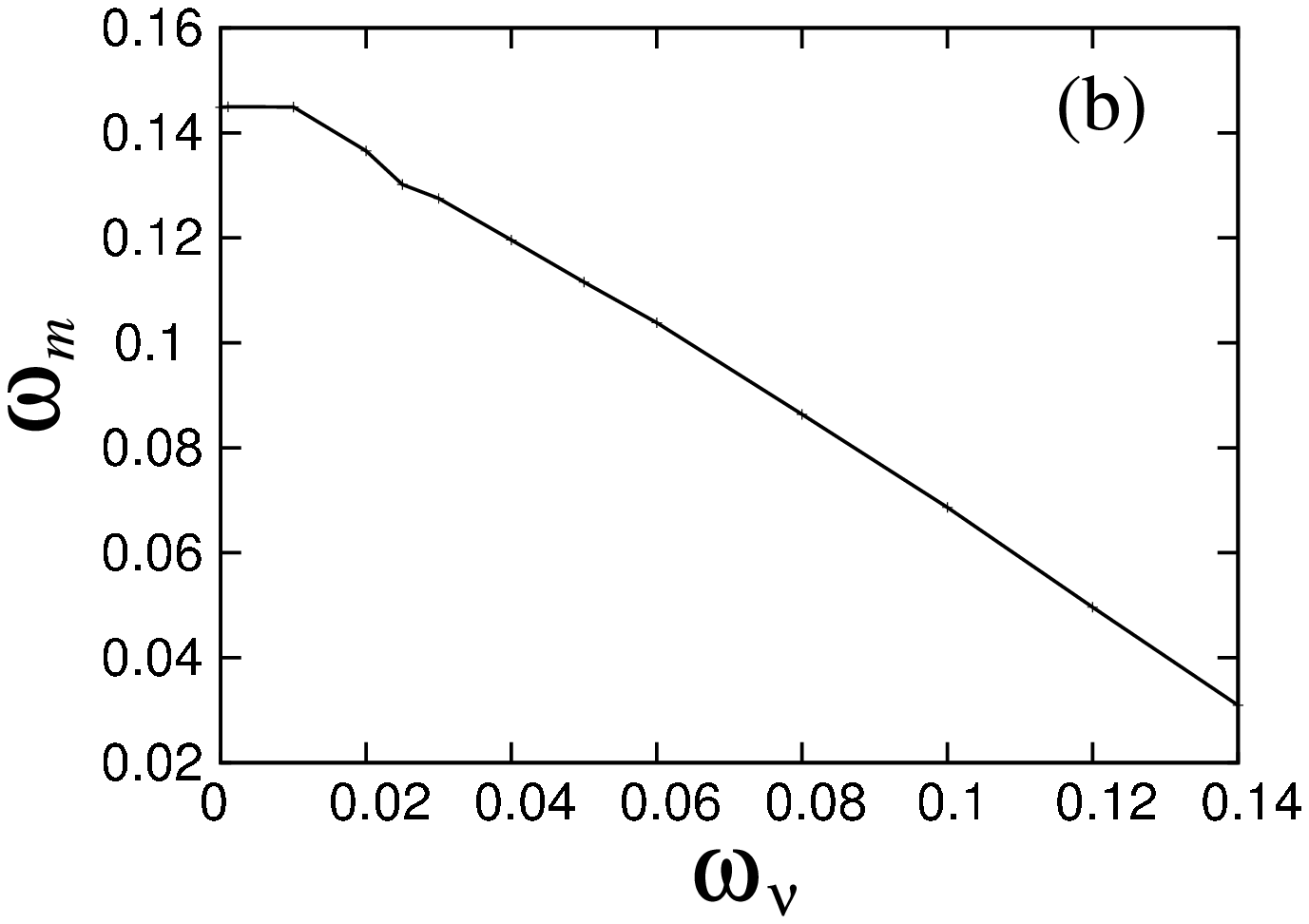} \\
& \\
\includegraphics[width=7.5cm] {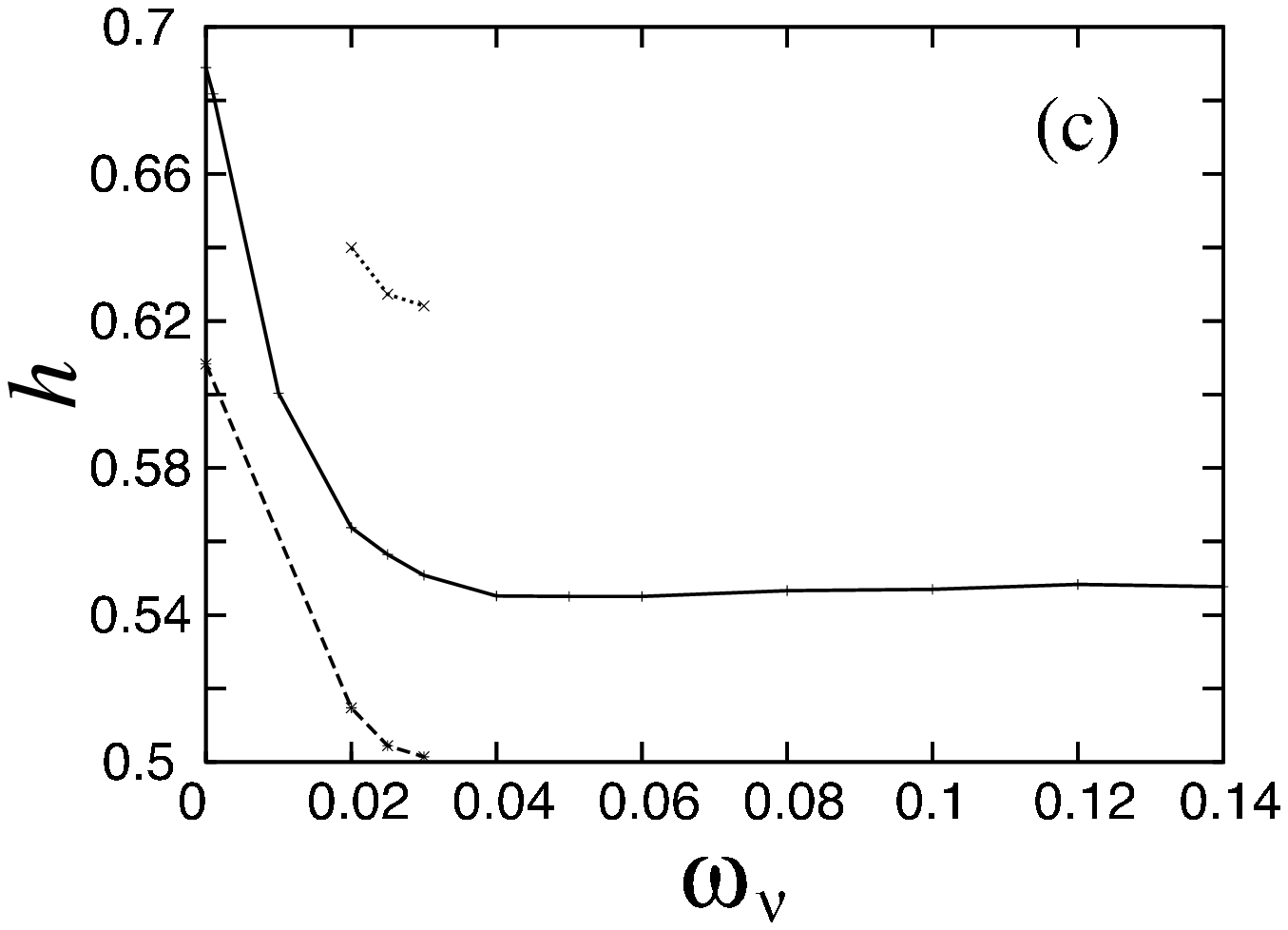} &
\hspace{0.4cm}
\begin{minipage}{7.5cm}
\vspace{-6.2cm}
\includegraphics[width=7.5cm] {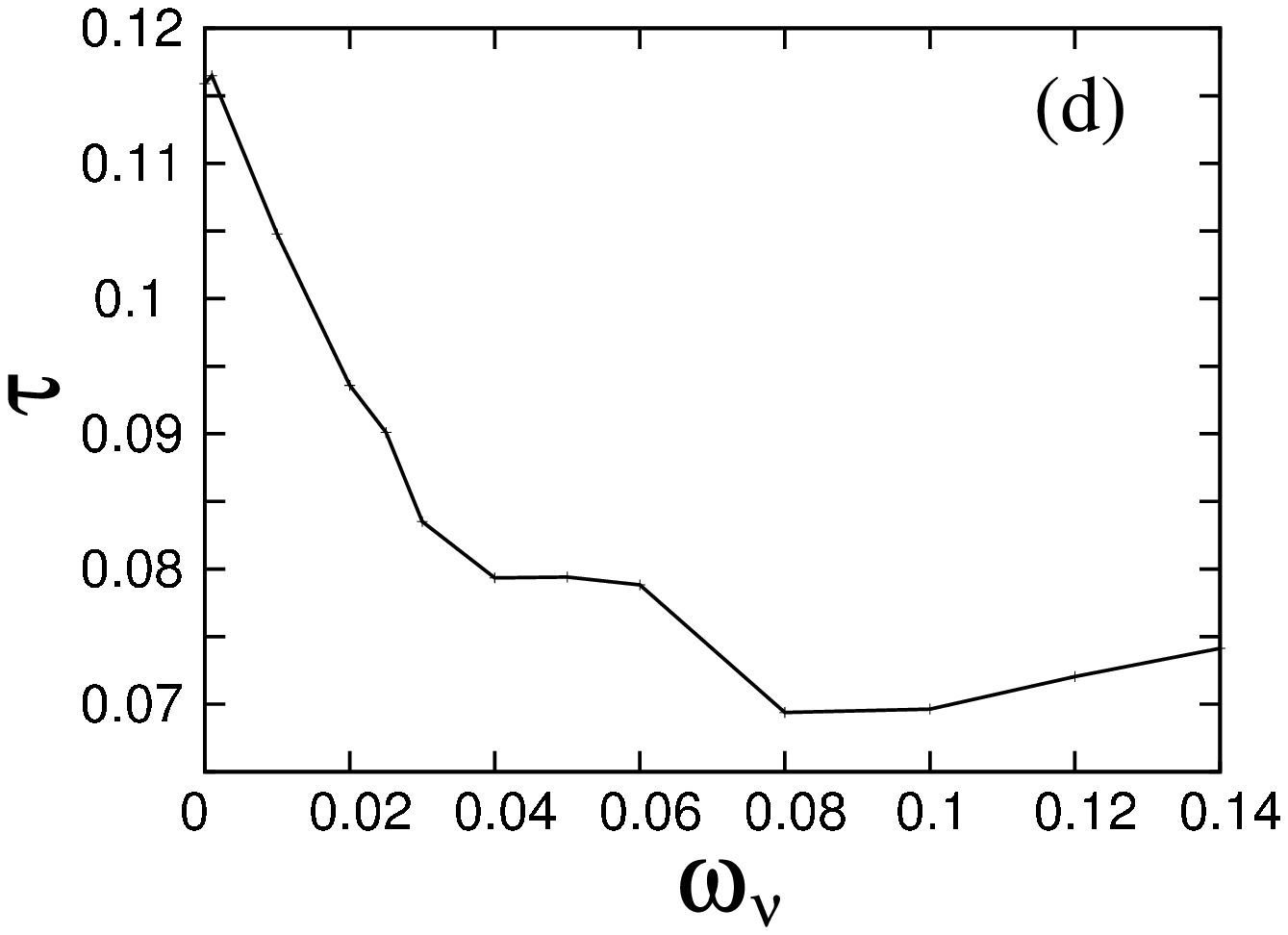}
\end{minipage} \\
\includegraphics[width=7.5cm] {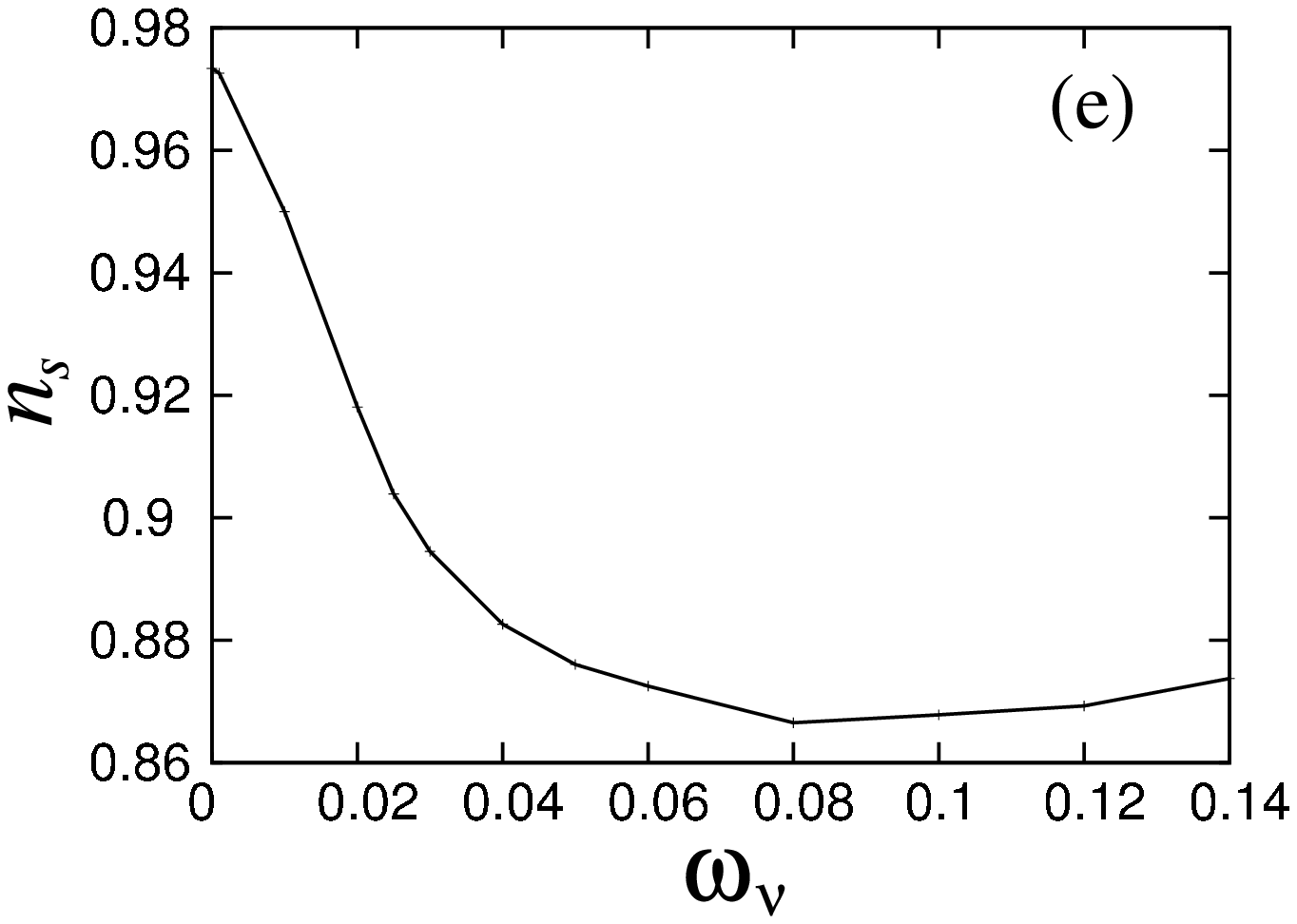} &
\hspace{0.5cm}\includegraphics[width=7.5cm] {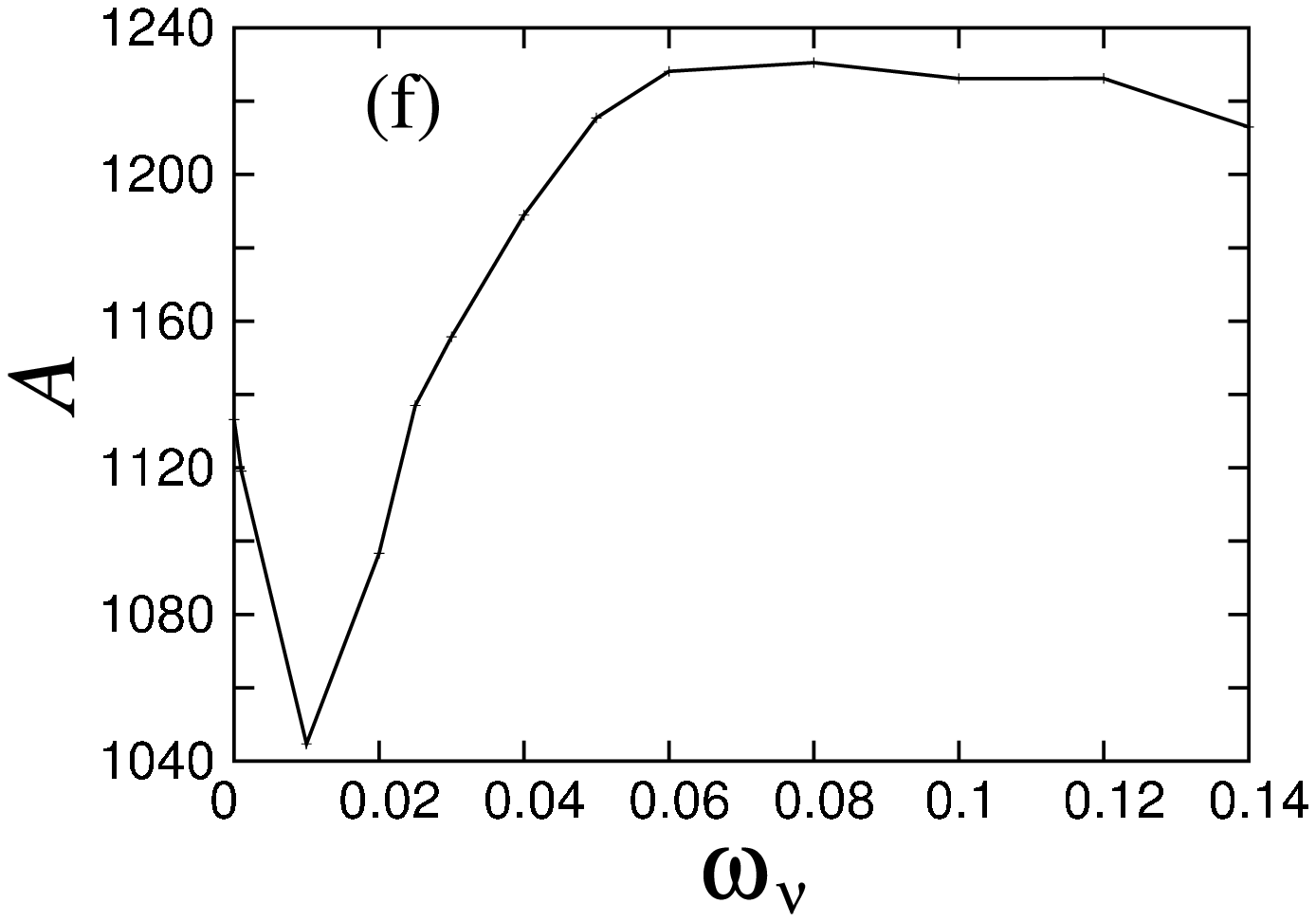} \\
& \\
\end{tabular}
\caption{The cosmological parameters for the solutions
that give minimum $\chi^2$ as a function of $\omega_{\nu}$.
The two line segmants shown in panel (c) are the cases
for a negative (dotted line) and a positive (dashed line)
curvature universe.
}
\label{fig:parameters_min}
\end{center}
\end{figure}

\begin{figure}
\begin{center}
\begin{tabular}{cc}	
\includegraphics[width=7.5cm]{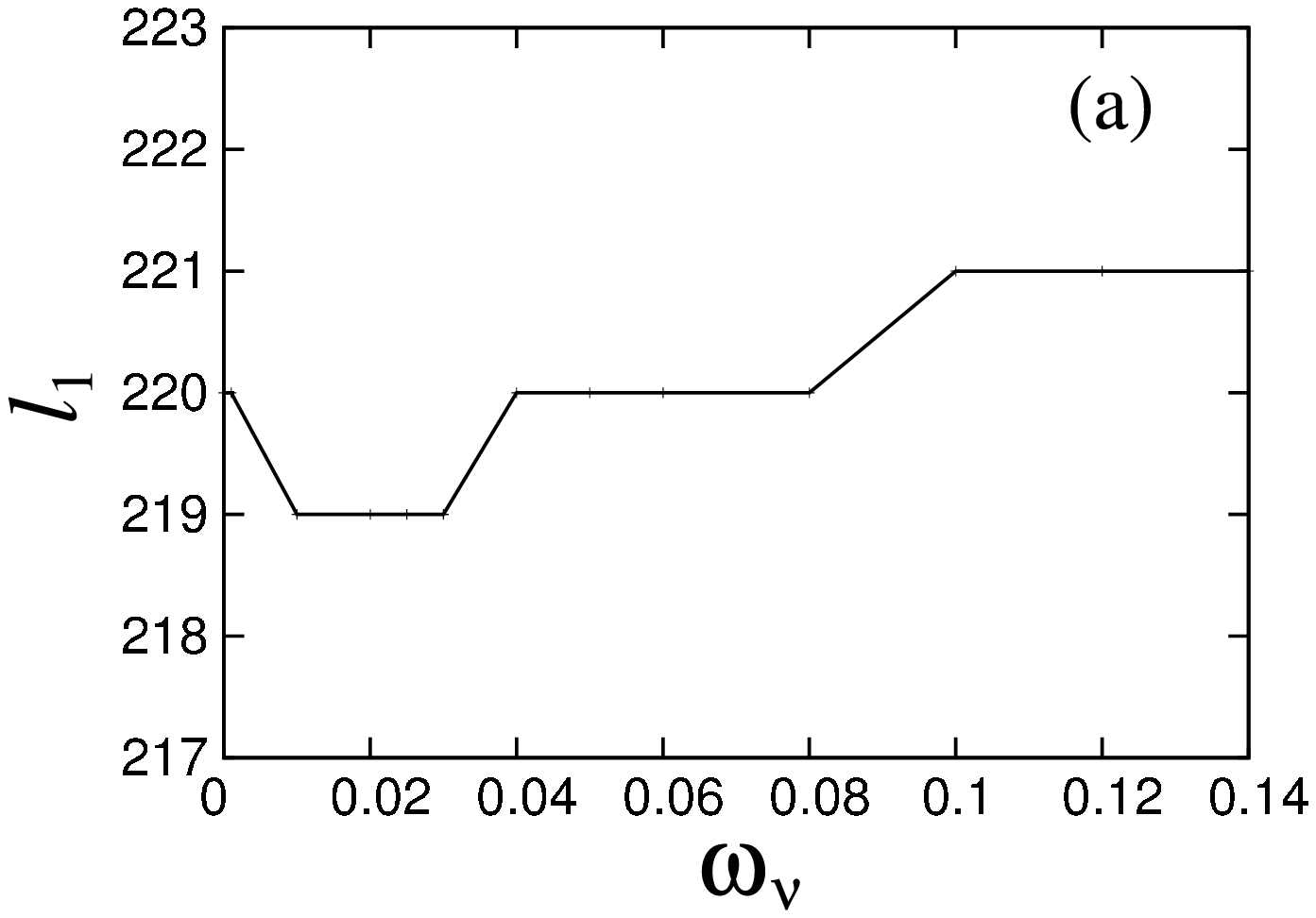} &
\hspace{0.5cm} \includegraphics[width=7.5cm]{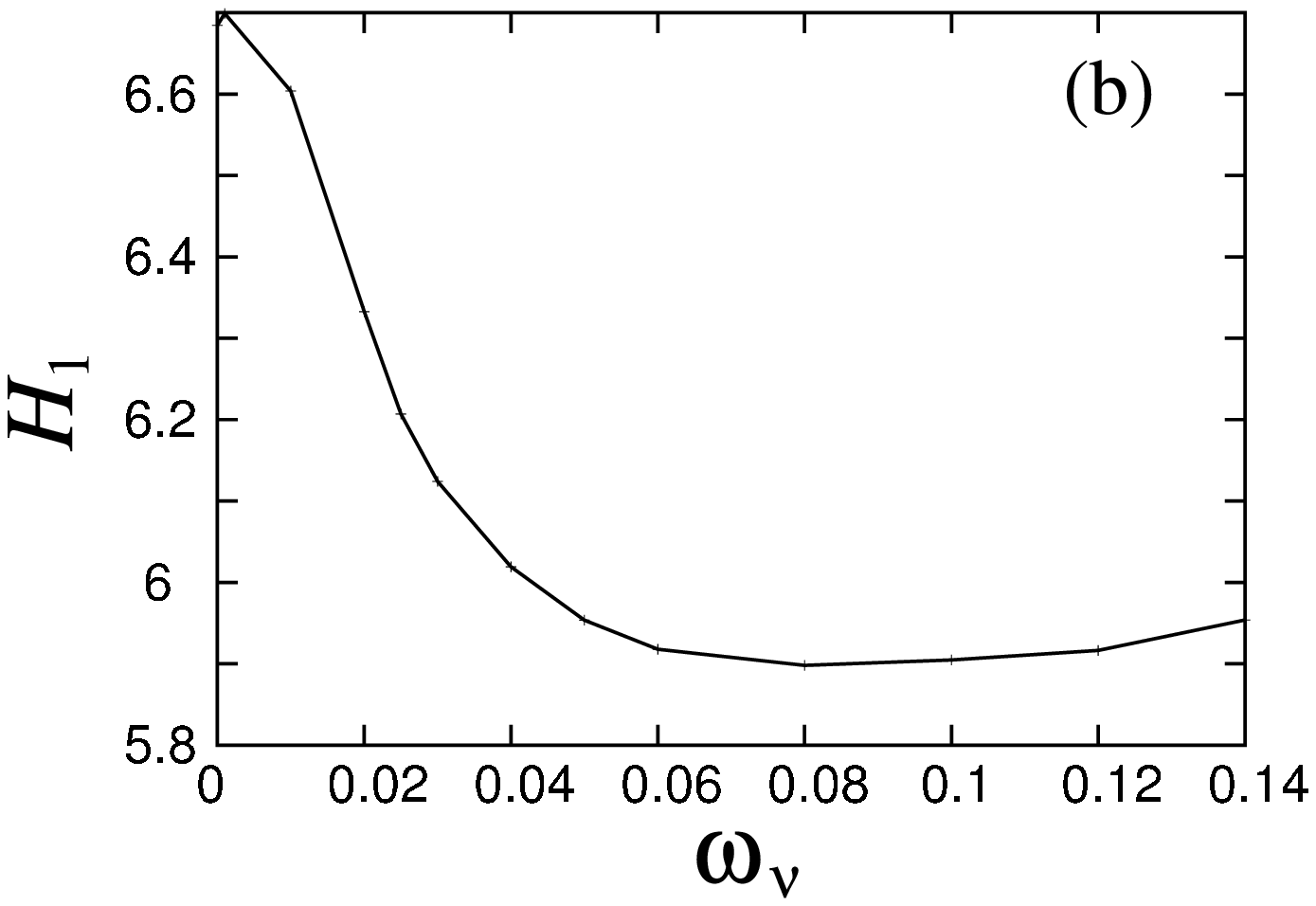} \\
& \\
\includegraphics[width=7.5cm]{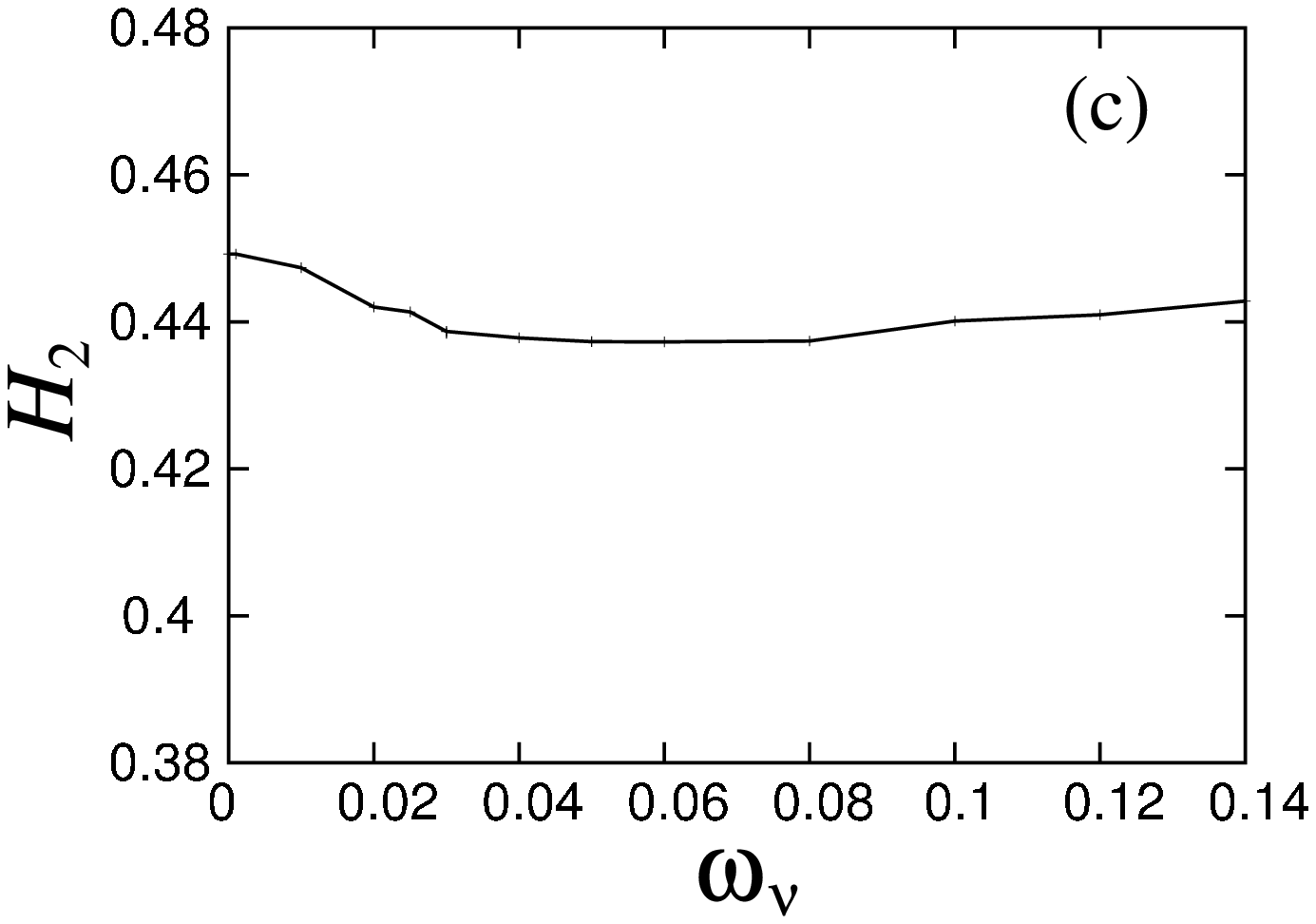} &
\hspace{0.5cm} \includegraphics[width=7.5cm]{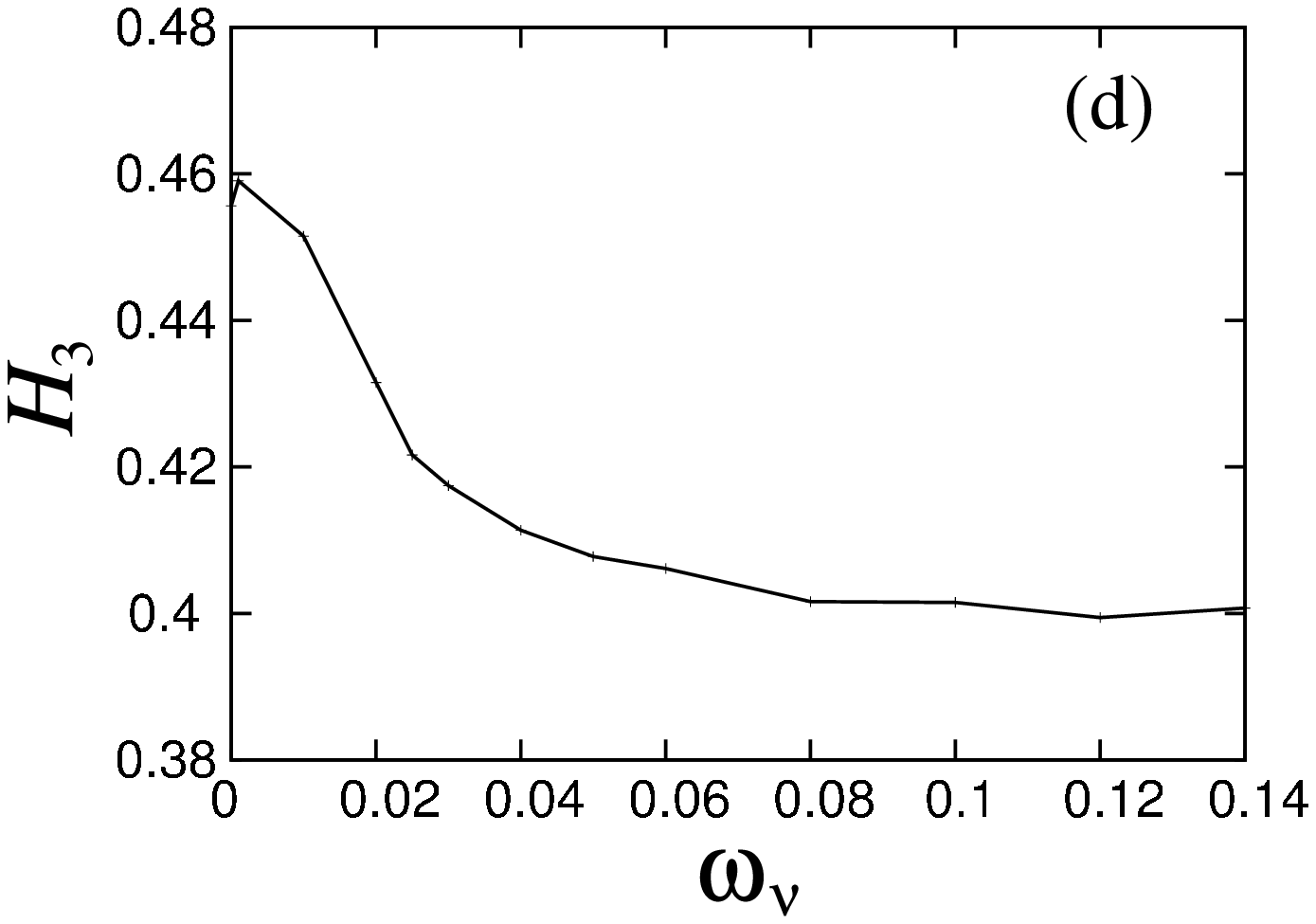} \\
& \\
\end{tabular}
\caption{The values of reduced CMB observables for the solutions
that give minimum $\chi^2$ as a function of $\omega_{\nu}$.
}
\label{fig:observables_min}
\end{center}
\end{figure}

The parameters of the $\Lambda$CDM model we shall consider
are the baryon density $\omega_b \equiv \Omega_b h^2$, the matter
density $\omega_m \equiv \Omega_m h^2$ (which includes baryons but
excludes neutrinos), the Hubble constant $h$, the
reionisation optical depth $\tau$, the scalar spectral index $n_s$ of
the power-law adiabatic perturbations, and overall
normalization $A$, where $\Omega_i$ denotes the energy density in units 
of
the critical density and $h$ is
$H_0 = 100\ h\ {\rm km}\ {\rm s}^{-1}{\rm
Mpc}^{-1}$. We ignore the tensor perturbations.
We define the normalisation parameter by
$A=\ell(\ell+1)C^{TT}_\ell/2\pi$ in units of $\mu$K$^2$ at $\ell=2$,
which differs from the WMAP definition.
In addition, we include the neutrino mass
density $\omega_{\nu}\equiv \Omega_{\nu} h^2$, which
is related to the neutrino mass as
\begin{equation}
\omega_{\nu} = \frac{\Sigma m_{\nu}}{94.1\ {\rm eV}}\ . \label{eq:omeganu}
\end{equation}
We assume three generations of neutrinos with their masses being
degenerate, $m_{\nu_e}= m_{\nu_{\mu}}= m_{\nu_{\tau}}$, so $m_{\nu} =
31.4\ \omega_{\nu}$ eV.
The vacuum energy is taken to satisfy the flat curvature
$\Omega_{\rm tot}\equiv \Omega_\Lambda+\Omega_m+\Omega_\nu=1$,
but this condition is relaxed
in Sec.~\ref{sec:nonflat}. We often write
$\omega_{\Lambda}\equiv\Omega_{\Lambda}h^2$.
We run CMBFAST \cite{Seljak:1996is} to calculate CMB multipoles
for the total of $1\times 10^6$ sets of parameters in the course
of our work.
The $\chi^2$ are computed for the entire temperature (TT)
and polarisation (TE) data set of WMAP (899 and 449 points, respectively) 
using the likelihood code supplied by the
WMAP team \cite{Verde:2003ey,Hinshaw:2003ex,Kogut:2003et}.

We search for the $\chi^2$ minimum for fixed $\omega_\nu$, and refer
to the resulting $\chi^2$ minimum for a fixed $\omega_\nu$ as
$\chi^2_{\rm min}(\omega_\nu)$.
We prefer to use a deterministic search for the minimum rather than
the Markov chain Monte Carlo (MCMC)
method that is popular in the recent work
for the CMB analysis, since we find that the latter, while it
is in principle efficient to find a gross global structure of the likelihood function,
often fails to yield the accurate shape of
the likelihood function away from the minimum unless the chain is
long. 

To search for the $\chi^2$ minimum in 6 parameter
($\omega_b$, $\omega_m$, $h$, $\tau$, $n_s$, $A$)
space, we adopt a nested grid search. Technically, we apply 
the Brent method \cite{brent} of the successive
parabolic interpolation to find a minimum with respect to one specific
parameter with other parameters
at a given grid, and successively apply this method
to remaining parameters to find the global minimum\footnote{
The initial range of the parameters we searched is wide, 
e.g., $0<h<1$ and $\Omega_\Lambda>0$ etc.  Note that the priors
do not play any important role in our grid search, unlike in MCMC
where the priors are crucial. Should we find the parameter region near 
the boundary that results in a meaningfully small $\chi^2$ and 
contributes non-negligibly to the likelihood function, we would simply
enlarge the parameter region for search. 
This never happened in our case, however.}. 
We describe more details of this minimisation procedure
in Appendix A.
If more than one conspicuous  $\chi^2$ minima are detected in
the process, we apply this method to each local minimum.
We run the CMBFAST code typically $10^5$ times to find the global 
minimum
for a given $\omega_\nu$. Note that the adoption of the
Brent method greatly reduces the number of
grids needed for a required accuracy.

In order to obtain the likelihood function with respect to a specifc
parameter, we must 
in principle integrate over the parameters other than the one that
concerns us. The $\chi^2$ function could be different significantly
from the true likelihood function, if the distribution is not Gaussian.
To verify this point,
we carry out an adaptive Monte Carlo integral using the Vegas
code \cite{lepage} to check if the likelihood function inferred from
the $\chi^2$ function differs significantly from that obtained by
integarting over parameter space.
The integral is performed 
for the cases of $\omega_\nu=0$ and 0.08, the latter
being the value with which Tegmark et al. give a rather high
likelihood. In particular, we want to check if a local minimum that gives
a relatively large $\chi^2$ is favoured from a large measure of
parameter space. In so far as we have examined, there is no evidence
that the likelihood inferred
from $\chi^2$ function differs significantly from that obtained from 
the integral (examples are shown below).  
In particular, we do not find the case where the integration measure 
overcomes an
excess  $\chi^2$: the parameter sets that give the global $\chi^2$ minimum
always represent the maximally likely parameters
in the case we studied.

\begin{table}
\begin{tabular}{|c|| c|c|c|c |c||c|}
\hline
  & $\omega_b$ & $\omega_m$ &   $h$ & $\tau$ & $n_s$   & $\chi^2_{min}$  
\\
\hline
Ours & 0.0230	& 0.145 &			0.689 &	0.116 &	0.973 &1428.6 \\	
\hline
Spergel et al. \cite{Spergel:2003cb} & 0.024$\pm$0.001 &0.14$\pm$0.02& 
0.72$\pm$0.05 & $0.166^{+0.076}_{-0.071}$ &	0.99$\pm$ 0.04 &1431  \\	
\hline
Tegmark et al. \cite{Tegmark:2003ud} & $0.0245^{+0.0050}_{-0.0019}$ & 
$0.140^{+0.020}_{-0.018}$	 &	$0.74^{+0.18}_{-0.07}$ 
&	$0.21^{+0.24}_{-0.11}$ &	$1.02^{+0.16}_{-0.06}$ &1431.5  \\	
\hline
\end{tabular}
\caption{Comparison of the solution for the massless neutrino with
those given by Spergel et al. and Tegmark et al. The errors stand for
one $\sigma$ confidence level.}
\label{tab:comparison}
\end{table}

\begin{table}
\begin{tabular}{|c|c|c|c |c|c||c||c|c |c|c|c|}
\hline
$\omega_b$ & $\omega_m$ &   $h$ & $\tau$ & $n_s$ & $A$
& $\chi^2$ &  $\ell_1$ &  $H_1$ &  $H_2$ & $H_3$ & remarks\\
\hline
0.0230	& 0.145 &			0.689 &	0.116 &	0.973 &	1133.1 &1428.6 &	
	220 &	6.68 &	0.449 &		0.456 & global mininum\\	
\hline
0.0305 &	0.121 &			0.957 &	0.487 &	1.21 &	1428.1 &	1428.8
&		221 &		6.31 &		0.453 &		0.481 &
local minimum\\	
\hline
\end{tabular}
\caption{Parameters for the two $\chi^2$ minima for 
$\omega_{\nu}=0$.}
\label{tab:two_chi2min}
\end{table}

The solutions that give a $\chi^2$ minimum for a given $\omega_\nu$
are presented in  Table~\ref{tab:chi2min}. The $\chi^2_{\rm 
min}$
as a function of the neutrino mass density is shown in
Figure~\ref{fig:chi2min}, and
the 6 parameters of the solution
for each neutrino mass density, also
given in the table, are displayed in Figure 2.
The corresponding four reduced CMB observables (defined below)
are shown in Figure 3 for the use in the next section. 

We first note that
the 6 parameters for $\omega_{\nu}=0$ agree with those
of the comparable solutions of
Spergel et al. \cite{Spergel:2003cb} and
Tegmark et al. \cite{Tegmark:2003ud}  within one
sigma errors (see Table \ref{tab:comparison}), verifying that our 
minimisation to find the global minimum works
at least as good as the MCMC method they used.
In fact, the overall $\chi^2$ we attained is appreciably smaller
than the two authors' for the same set of input data
($\chi^2_{\rm spergel}-\chi^2_{\rm ours}=2.4$, and
$\chi^2_{\rm tegmark}-\chi^2_{\rm ours}=2.9$). We may ascribe this to
a finer grid of the parameters
close to the minimum in our work.
We find bimodal structure of the
$\chi^2$ surface, most clearly visible for 
$n_s$ and $\tau$ that are strongly correlated to each other 
\cite{Tegmark:2003ud}. 
The two minima are found at $n_s=0.973$ and $n_s=1.21$
with the second minimum having a slightly larger $\chi^2$,
$\chi^2(n_s=1.21)-\chi^2(n_s=0.973)=0.2$, or the relative likelihood of 1.1: see Table \ref{tab:two_chi2min}.
The two parameter sets are disjoint by
a hill with a height more than one $\sigma$. 
The Vegas integration over multiparameter space centred on the two extrema 
indicates that the former minimum is favoured over the
latter by the ratio of 1.3 in terms of the likelihood value.
That is, likelihood from the $\chi^2$ estimator is a good approximation
to the `true' value obtained by marginalising the parameters, i.e.,
even in this case where the distribution is deviated from Gaussian
the $\chi^2$ function
is likely a reasonable approximation of the likelihood function.
Furthermore, we observed that the one-parameter distributions with respect to the other five parameters are close 
to Gaussian once we require $n_s$ to be around the peak at
$n_s=0.973$ (see Appendix B). 
This suggests that the distribution in multidimensional space is likely not far from the Gaussian.  Hence, we infer that the $\chi^2$ statistics
well approximates the reality.

The bimodal structure we find is
consistent with what was found by Tegmark et al., but our
likelihood of the second minimum is much higher than
that reported (the ratio of likelihoods between the two extrema 
by Tegmark et al.~is 2.5). 
We suspect that Markov chain of Tegmark et al.
does not sample well around the second minimum. This point is
demonstrated in more detail in Appendix B. This is an example that
the current application of the MCMC does not give an accurate
likelihood function away from the global minimum. 
Of course, the second solution is an unphysical
one in the sense that it is allowed only at the cost of an unacceptably high
reionisation optical depth ($\tau\approx 0.5$); the solution is
deleted with some prior on $\tau$. The resulting parameters 
$\omega_b$, and $h$ are also deviated significantly from the 
values derived from other observations.

In Figure \ref{fig:chi2min} we observe that
$\chi^2_{\rm min}(\omega_{\nu})$ increases with the neutrino mass density.
The curve of $\chi^2$ minimum is close to a parabola
except in the immediate vicinity of $\omega_\nu=0$. Taking 
$\Delta\chi^2 \equiv \chi^2_{\rm min}(\omega_{\nu}) - \chi^2_{\rm min} = 4$
to set an upper limit on
$\omega_{\nu}$ at the 95\% confidence level,  we obtain
\begin{eqnarray}
\omega_{\nu} < 0.024, \hskip5mm{\rm or}\hskip5mm m_{\nu} < 0.75\ {\rm eV}.
\label{eq:limit}
\end{eqnarray}

Since the likelihood function with respect to $n_s$ and $\omega_\nu$, ${\cal L} = \exp[-\Delta \chi^2(n_s,\omega_\nu)/2]$, which is constructed by minimising the five other parameters, is visibly deviated from Gaussian, we integrated it over $n_s$ and then over $\omega_\nu$. This yields the 95\% confidence limit
\begin{eqnarray}
\omega_{\nu} < 0.021, \hskip5mm{\rm or}\hskip5mm
m_{\nu}<0.66\ {\rm eV},
\end{eqnarray}
which is close 
to Eq.~(\ref{eq:limit}), a  simple reading from $\chi^2$. 
[The difference primarily comes from the second peak of the 
$\chi^2$ function, which is ignored in Eq.~(\ref{eq:limit})].
If the distributions of the five other
parameters are close to Gaussian, a two-dimensional integral is sufficient
to obtain an accurate likelihood.

We cannot compare this limit on the neutrino mass directly with those
derived in Spergel et al. and Tegmark et al. \cite{Tegmark:2003ud},
in which those authors used the galaxy clustering data as
additional inputs.
On the other hand, the latter authors claim that
WMAP alone does not give a limit on the neutrino mass and that the
massive neutrinos
can make up 100\% of dark matter at about one $\sigma$ confidence
unless galaxy clustering data are used.  Our result contradicts this.
We do not find a parameter set that gives acceptable $\chi^2$
for the neutrino mass density beyond the limit. Furthermore, the measure of
the parameter space does not seem to increase for a larger $\omega_\nu$.
Our Vegas integrals give a relative likelihood between $\omega_\nu=0$
and $\omega_\nu=0.08$ to be $7\times 10^{-5}$, which is consistent
with the estimate from our $\chi^2$ curve $5\times 10^{-5}$, whereas
Tegmark et al.'s value is 0.6.
We suspect that sampling of the Markov chain of Tegmark et al. does not  
give an accurate likelihood function away from $m_\nu=0$ that is
the global minimum, as similarly happened with the case of 
$n_s$ discussed above and in Appendix B.
In particular, we do not find a mixed-dark-matter-model 
($\Omega_m+\Omega_\nu=1$) like solution:
the CMB multipoles of the hot dark matter
model with some sets of parameters are visibly similar
to the observation \cite{Elgaroy:2003yh}, but a closer inspection shows that
$\chi^2$ is always 
unacceptably
large, given a high accuracy of the WMAP data\footnote{
For the set of parameters of a mixed-dark-matter-model like solution
proposed by Elgar\o y \& Lahav \cite{Elgaroy:2003yh}, we find
$\chi^2=1482$, which is larger than that of the 
$\Lambda$CDM solution by $\Delta \chi^2=50$. 
We cannot make $\chi^2$ significantly smaller around this solution.}.
In the following
section we see a reason how can one obtain the limit on the neutrino
mass density from the CMB data alone.

We remark that the current WMAP TE data do not seem to play a significant role
in deriving our limit, as we find in separate runs of the $\chi^2$
minimisation using only the TT data\footnote{
We somewhat loosened the convergence criteria for these runs, but
we still obtained $\chi^2_{\rm min}=972.3$ compared with 972.4 of 
Tegmark et al. The solution differs appreciably from that with the full data
set only in $\tau$, which for the TT case is close to zero.}:
the $\chi^2$ curves differ little between the two cases.
This somewhat differs from the forecast of 
Eisenstein et al.\cite{Eisenstein:1998hr}
who indicated a tighter error allowance that would result with the
WMAP polarisation data\footnote{Their forecast 2 $\sigma$ errors 
are 1.2 eV with the
polarisation data, and 1.8 eV without them for a hypothetical neutrino
mass of 0.7 eV assuming idealised CMB data of the Gaussian variance
around the prediction of the $\Lambda$CDM model. 
This does not contradict our actual limit.}.

As a final remark, the two $\chi^2$ minima found for $\omega_{\nu}=0$
persist up to $\omega_{\nu} \sim 0.04$, but the one that 
corresponds to the ``unphysical solution'' disappears 
for $\omega_{\nu} \gtrsim 0.05$.

\section{The reduced CMB observables and the neutrino mass}
\label{sec:reason}

\subsection{The reduced CMB observables and the goodness of the 
$\Lambda$CDM fit}
Following  Ref.~\cite{Hu:2000ti}, we focus on four quantities which
characterise the shape of the CMB spectrum: the position of the first
peak $\ell_1$, the height of the first peak relative to the large 
angular-scale
amplitude evaluated at
$\ell=10$,
\begin{equation}
H_1 \equiv \left( \frac{\Delta T_{l_1}}{\Delta T_{10}} \right)^2
\label{eq:H1},
\end{equation}
the ratio of the second peak height to the first,
\begin{equation}
H_2 \equiv \left( \frac{\Delta T_{l_2}}{\Delta T_{l_1}} \right)^2
\label{eq:H2},
\end{equation}
and the ratio of the third peak height to the first,
\begin{equation}
H_3 \equiv \left( \frac{\Delta T_{l_3}}{\Delta T_{l_1}} \right)^2
\label{eq:H3},
\end{equation}
where $(\Delta T_l)^2 \equiv l(l+1)C^{TT}_l/2\pi$ and $C^{TT}_l$ is the
multipole coefficient of the temperature anisotropy.

Taking the advantage that we generated one million CMB templets,
we estimate the reduced CMB observables from the envelope drawn by
the entire set of the templets.
Our sampling is dense enough to define the correct envelope at least for small $\Delta\chi^2$ that concerns us.
 The result is
\begin{eqnarray}
\ell_1 &=& 220^ {+1.5}_{-1} \label{eq:error_l1}, \\
H_1 &=& 6.7^{+0.3}_{-0.6} \label{eq:error_H1}, \\
H_2 &=& 0.449\pm0.007\label{eq:error_H2}, \\
H_3 &=& 0.46^{+0.04}_{-0.02} \label{eq:error_H3},
\end{eqnarray}
which is shown in Figure 4. 
The error is 1 standard deviation obtained by halving the range that
gives 2 $\sigma$ error, i.e.,
$\Delta \chi^2 \equiv \chi^2 - \chi^2_{\rm min} = 4$,
because the structure of the $\chi^2$
curve is not always parabolic at around $\Delta \chi^2\approx 1$.
The central values are the best fit solution given in Table 1.
Eqs.~(\ref{eq:error_H2}) and (\ref{eq:error_H3})
are consistent with the values Tegmark et al. \cite{Tegmark:2003ud}
quoted for their best parameter set ($H_1$ is not given). 
We note particularly small errors for $\ell_1$ and $H_2$, which play
an important role in the argument given in the next subsection.
In addition, we draw the envelopes for the case of a few non-zero 
neutrino
masses. They give increasingly larger $\chi^2$ as the mass increases,
in particular for $\omega_\nu\geq 0.02$;
the widths of the $\chi^2$ valleys become somewhat narrowed as
$\omega_\nu$ increases.

We also attempt to obtain the four reduced CMB observables 
from the fits that give a
$\chi^2$ minimum for a restricted range of $\ell$
   using our CMB templets,
as was done in \cite{Hu:2000ti}. We calculate $\chi^2$ using
the TT data of appropriate multipole ranges. 
We use $75 \leq l \leq 375$ for $\ell_1$, $7 \leq l \leq 375$ for
$H_1$, $75 \leq l \leq 375$ and $450 \leq l \leq 600$ for $H_2$, and
$75 \leq l \leq 375$ and $750 \leq l \leq 875$ for $H_3$. The results
are displayed in Figure~4 
above.
We obtain
\begin{eqnarray}
\ell_1 &=& [~219,~ 222~] \label{eq:error2_l1}, \\
H_1 &=& [~6.5,~ 7.9~] \label{eq:error2_H1}, \\
H_2 &=& [~0.430,~ 0.452~]\label{eq:error2_H2}, \\
H_3 &=& [~0.362,~ 0.488~] \label{eq:error2_H3}.
\end{eqnarray}

The numbers bracketed are the 1 $\sigma$ range obtained by halving
the 2 $\sigma$ range of the $\chi^2_{\rm local}$ 
curve\footnote{The 1$\sigma$ range of 
$H_1$ depends on the choice of the
lower limit of the $\ell$-range.
It is well known that $\ell=2$ and 3 multipoles are anomalously
low compared to the expectation from the $\Lambda$CDM model.
If the lower limit is set to $\ell_{\rm min}=2$, the one $\sigma$ range will be
$H_1=[7.0, 8.0]$. The 1 $\sigma$ range nearly converges for 
$\ell_{\rm min}\ge 3$:
the central value does not differ from Eq.~(\ref{eq:error2_H1})
more than 0.1.}.
An inspection of the fits of the templets to the obeserved CMB multipoles
indicates that the data are well represented by those templets.
Figure 4 may give the impression that 
the $\chi^2$ curves do not agree with those obtained from
the envelope of the global $\Lambda$CDM fit:
the valley of the $\chi^2$ curves is generally wider, and the positions of
the bottom of valley is somewhat shifted;
the $\chi^2_{\rm local}$ of the best global fit 
solution is larger than the best local
fits by $\Delta\chi^2\approx2$.
 The central values of
Eqs.~(\ref{eq:error_l1}) to (\ref{eq:error_H3}), however, fall
in the one $\sigma$ range of  Eqs.~(\ref{eq:error2_l1}) to
(\ref{eq:error2_H3})\footnote{
The parameters derived by Page et al.~\cite{Page:2003fa},
who extracted them by fitting the WMAP data by Gaussian
and parabolic functions,
$\ell_1=220.1\pm 0.8$, $H_2=0.426\pm 0.015$, and
$H_3=0.42\pm 0.08$  ($H_1$ is not given) deviate from our
$\Lambda$CDM solutions in Eqs.~(\ref{eq:error_l1}) to 
(\ref{eq:error_H3}) by up to 1.5$\sigma$,
but agree with those given in  Eqs.~(\ref{eq:error2_l1}) to
(\ref{eq:error2_H3}).}.
Our analysis, showing that
the best global fit and the local fits resulted in the consistent
reduced CMB parameters within 1 $\sigma$, leads us to conclude the
goodness of the $\Lambda$CDM fit.

For the consideration given in the next subsection, where we are
concerned with the problem how much massive neutrinos increase
$\chi^2$ for the CMB data relative to the $\omega_\nu=0$ solution,
we should use Eqs.~(\ref{eq:error2_l1}) - (\ref{eq:error2_H3}),
rather than Eqs.~(\ref{eq:error_l1}) - (\ref{eq:error_H3}),
which are obtained by restricted parameter searches.

\begin{figure}
\begin{center}
\begin{tabular}{cc}	
\includegraphics[width=7cm]{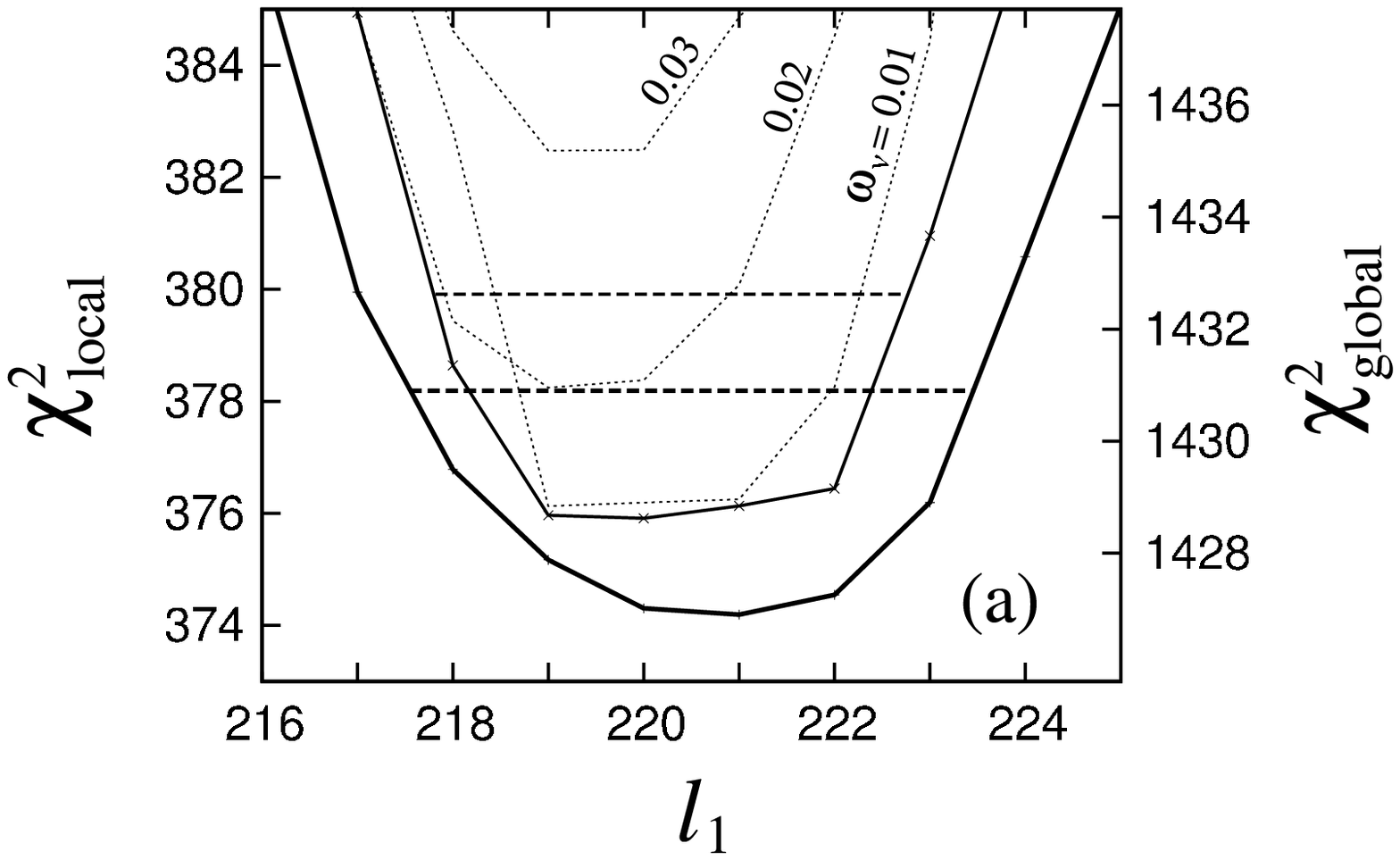} &
\hspace{2cm} \includegraphics[width=7cm]{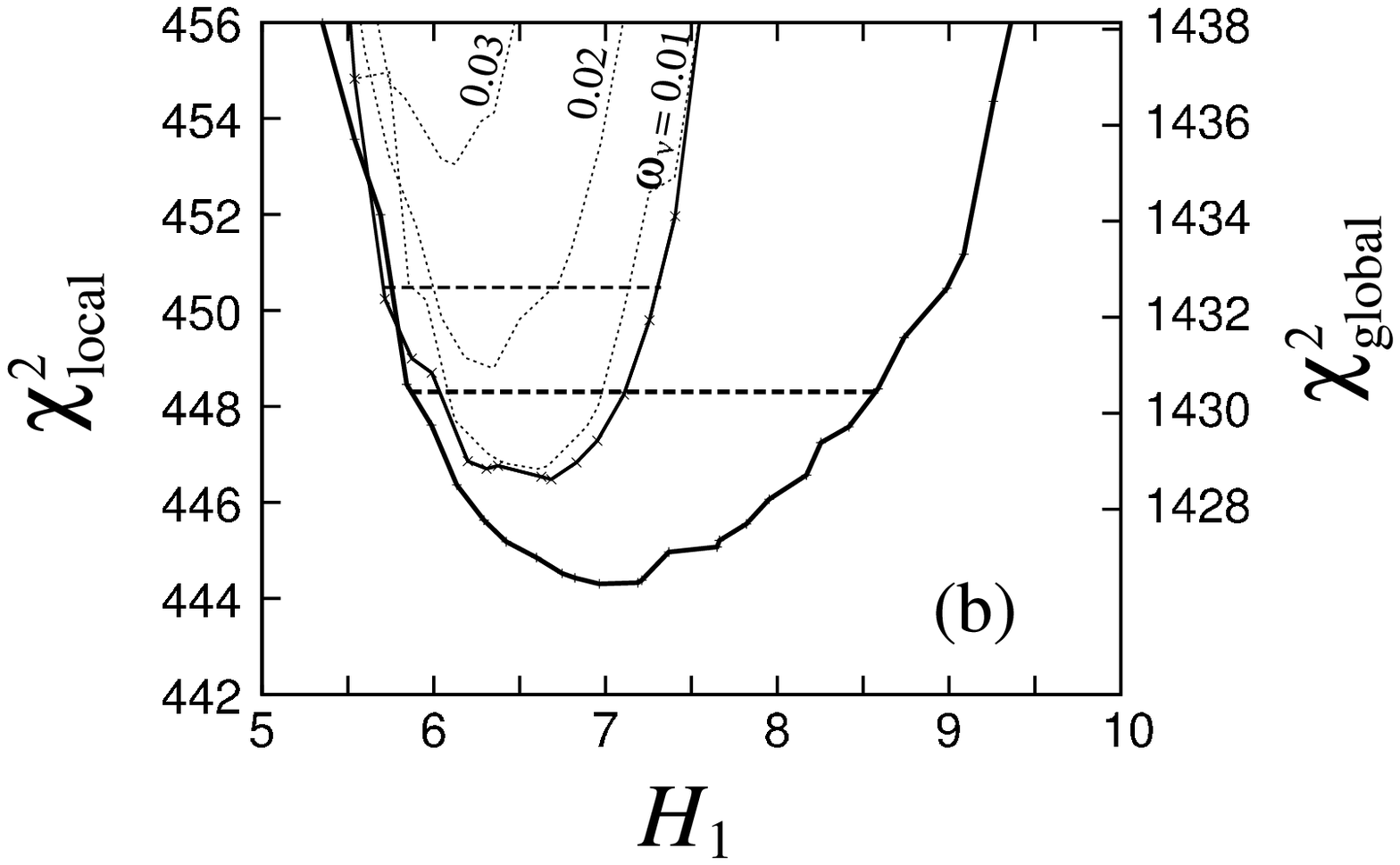} \\
& \\
\includegraphics[width=7cm]{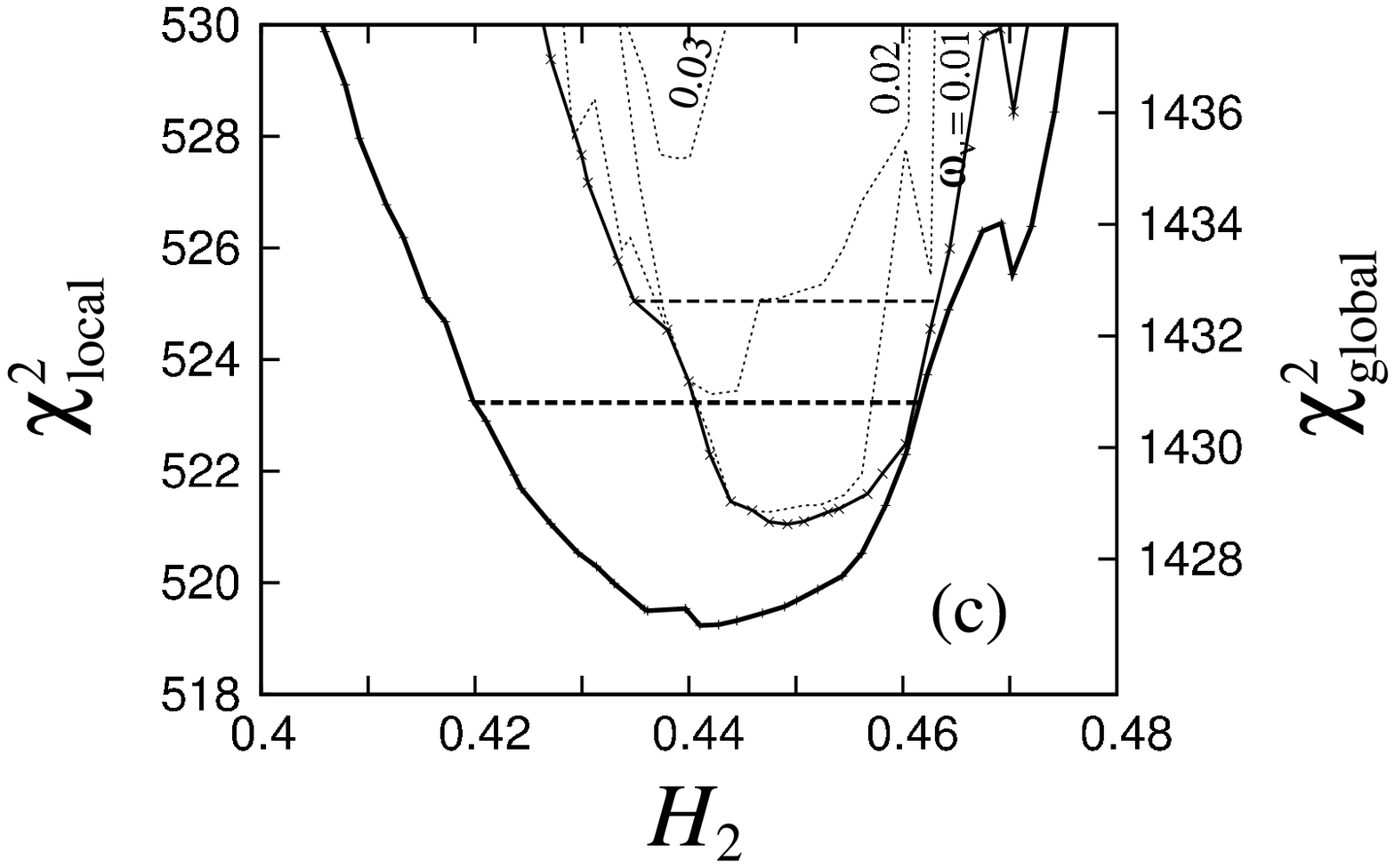} &
\hspace{2cm} \includegraphics[width=7cm]{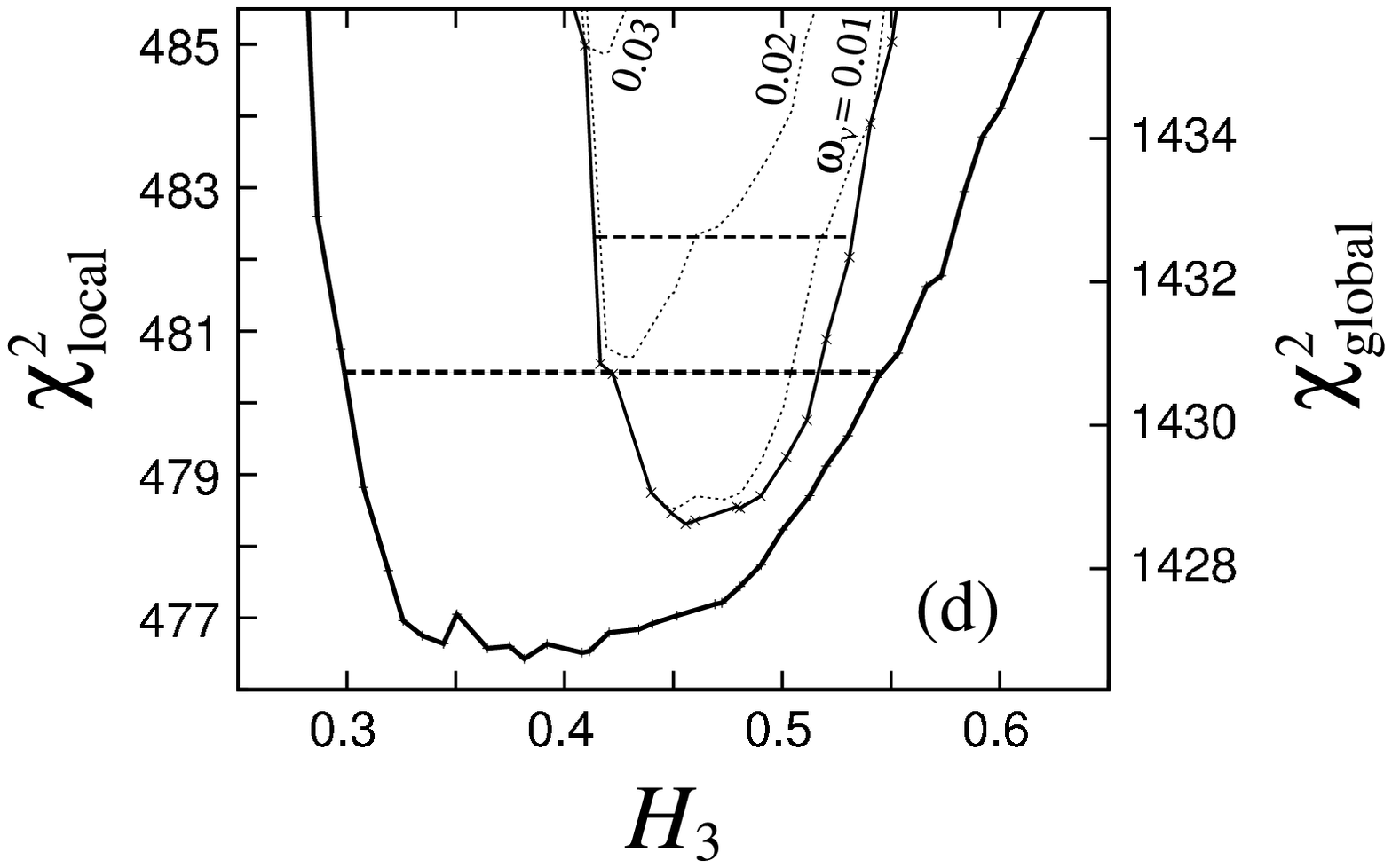} \\
& \\
\end{tabular}
\caption{Constraints on the four reduced CMB observables. Local 
$\chi^2$ is computed using restricted sets of multipoles as explained 
in the text and is measured with $\chi^2_{\rm local}$ in the relevant range
indicated in the left ordinate. The $\chi^2$ of global solution is
measured for the value with respect to the entire data set as 
measured in the right ordinate. The relative normalisation is 
fixed so that the global solution that gives $\chi^2$ minimum gives
the local $\chi^2$ value measured in the left ordinate.
Dotted curves are the envelopes for $\omega_\nu=0.01$, 0.02 and
0.03 in the order of increasing minimum $\chi^2$.
The horizontal dashed line segments show the position of
$\chi^2-\chi^2_{\rm min}=4$.
}
\label{fig:observables_uncertainty}
\end{center}
\end{figure}

\subsection{Reduced CMB observables and the neutrino mass}

We calculate the response of the observables $O_i=\ell_1$, $H_{1}$ $H_2$
and $H_{3}$
with respect to the variation of cosmological parameters $x_j$, i.e.,
the partial derivatives
$\partial O_i/\partial x_j$, around the global best fit,
following Ref.~\cite{Hu:2000ti}.
We vary the parameters typically by $\pm$50\%
with a step of 5\% and take the difference from the reference values.
We find that the responses are quite
linear against the amount of the variations of the 6 parameters.
The exception is the response to the neutrino density parameter,
for which it is shown separately.
The resulting partial derivatives are:
\begin{eqnarray}
\Delta \ell_1 &=& 16\frac{\Delta \omega_b}{\omega_b} -25\frac{\Delta
\omega_m}{\omega_m} -47\frac{\Delta h}{h} +36\frac{\Delta n_s}{n_s}
+f_{\Delta \ell_1}(\omega_\nu)
\label{eq:deriv_l1}, \\
\Delta H_1 &=& 3.0\frac{\Delta \omega_b}{\omega_b} -3.0\frac{\Delta
\omega_m}{\omega_m} -2.2\frac{\Delta h}{h}-1.7\frac{\Delta \tau}{\tau}
+18\frac{\Delta n_s}{n_s}
+f_{\Delta H_1}(\omega_\nu)
\label{eq:deriv_H1}, \\
\Delta H_2 &=& -0.30\frac{\Delta \omega_b}{\omega_b} +0.015\frac{\Delta
\omega_m}{\omega_m} +0.41\frac{\Delta n_s}{n_s}
+f_{\Delta H_2}(\omega_\nu)
\label{eq:deriv_H2}, \\
\Delta H_3 &=& -0.19\frac{\Delta \omega_b}{\omega_b} +0.21\frac{\Delta
\omega_m}{\omega_m} +0.56\frac{\Delta n_s}{n_s}
+f_{\Delta H_3}(\omega_\nu)
\label{eq:deriv_H3}.
\end{eqnarray}
Here $\Omega_{\rm tot}=1$ is kept fixed, and $f_{\Delta O_i}(\omega_\nu)$ 
stands for
the variation with respect to the neutrino mass density.
The responses of $H_2$ and $H_3$ to $h$, and those of
$\ell_1$, $H_2$ and $H_3$ to $\tau$ are small, so
they are omitted in the expressions.
Page et al.~\cite{Page:2003fa}, evaluated the partial derivatives
for $H_2$ and $H_3$ to the variations of $\omega_b$,
$\omega_m$ and $n_s$ for the WMAP data using the analytic expressions 
\cite{Hu:2000ti}. Our empirical derivatives for these quantities 
are
consistent with their analytic evaluation.

\begin{figure}
\begin{center}
\begin{tabular}{cc}	
\includegraphics[width=7.5cm]{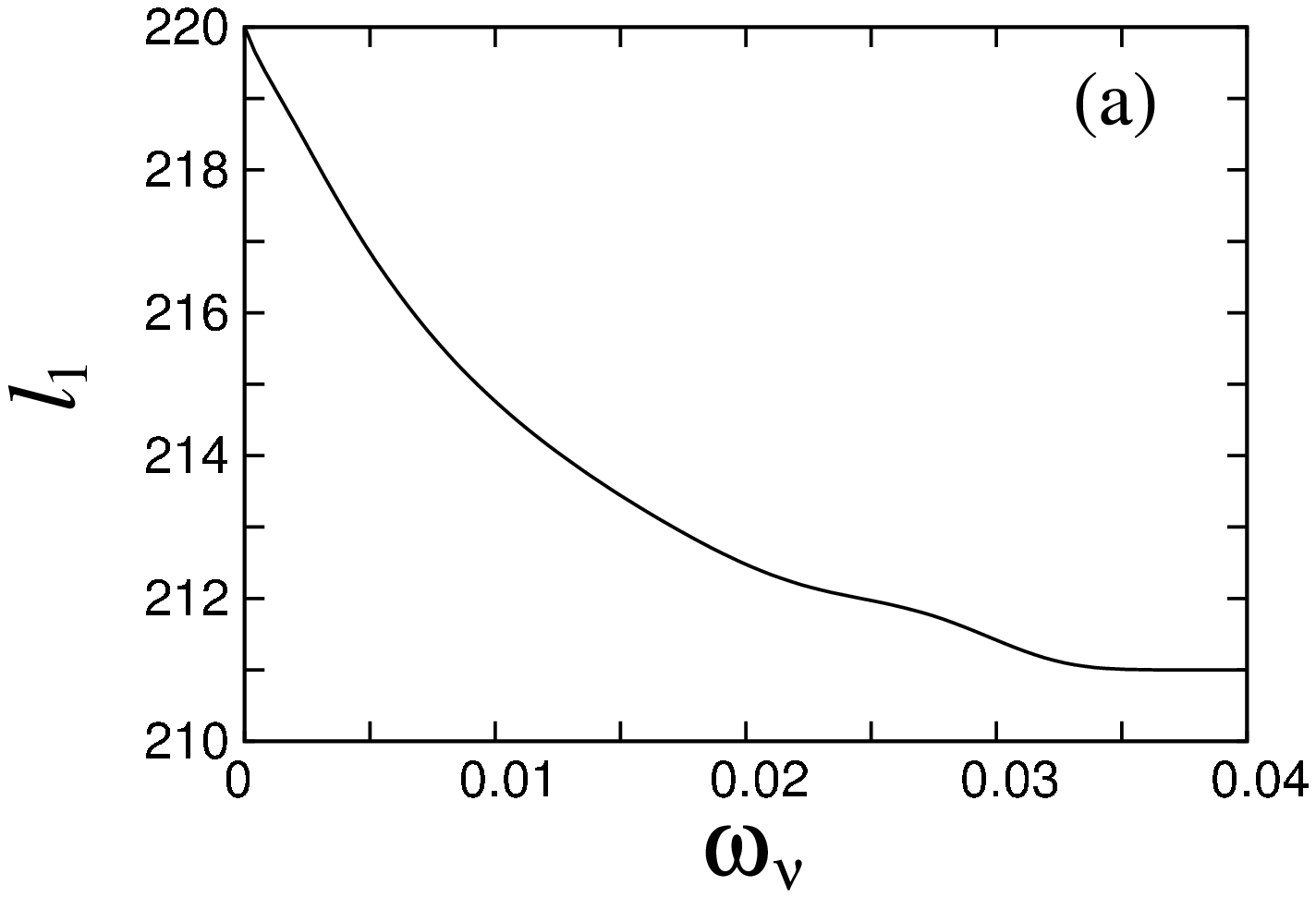} &
\hspace{0.5cm} \includegraphics[width=7.5cm]{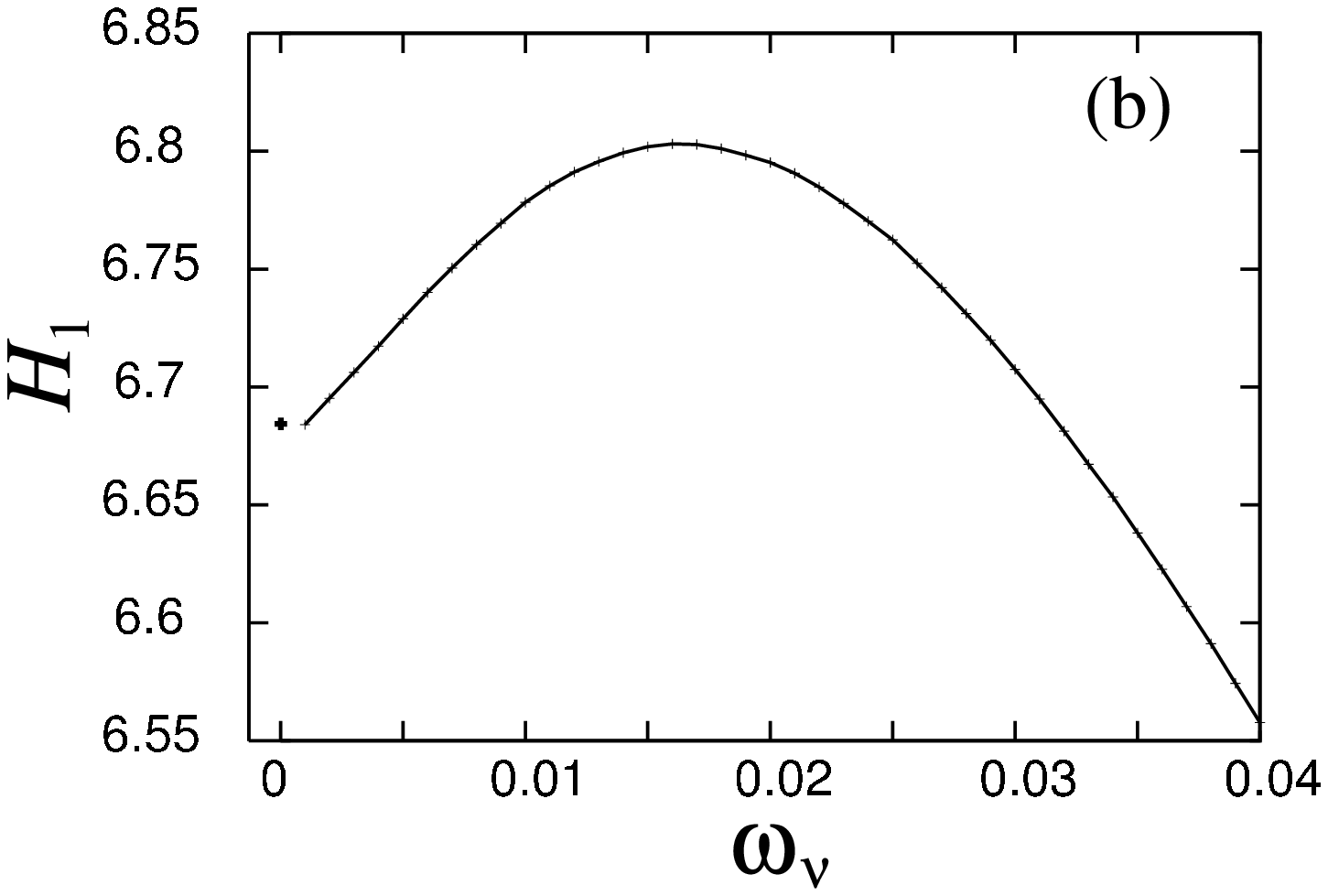} \\
 & \\
\includegraphics[width=7.5cm]{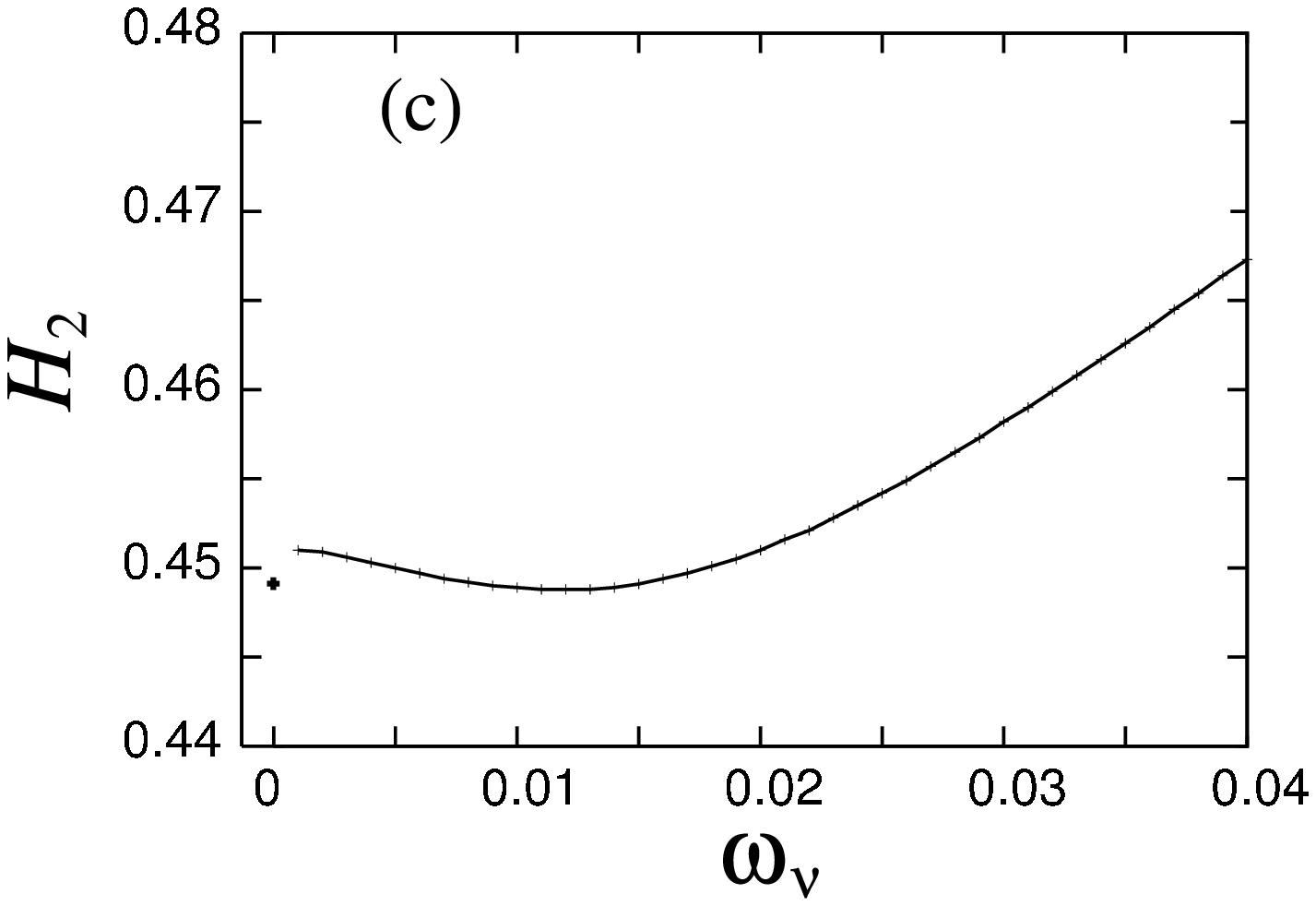} &
\hspace{0.5cm} \includegraphics[width=7.5cm]{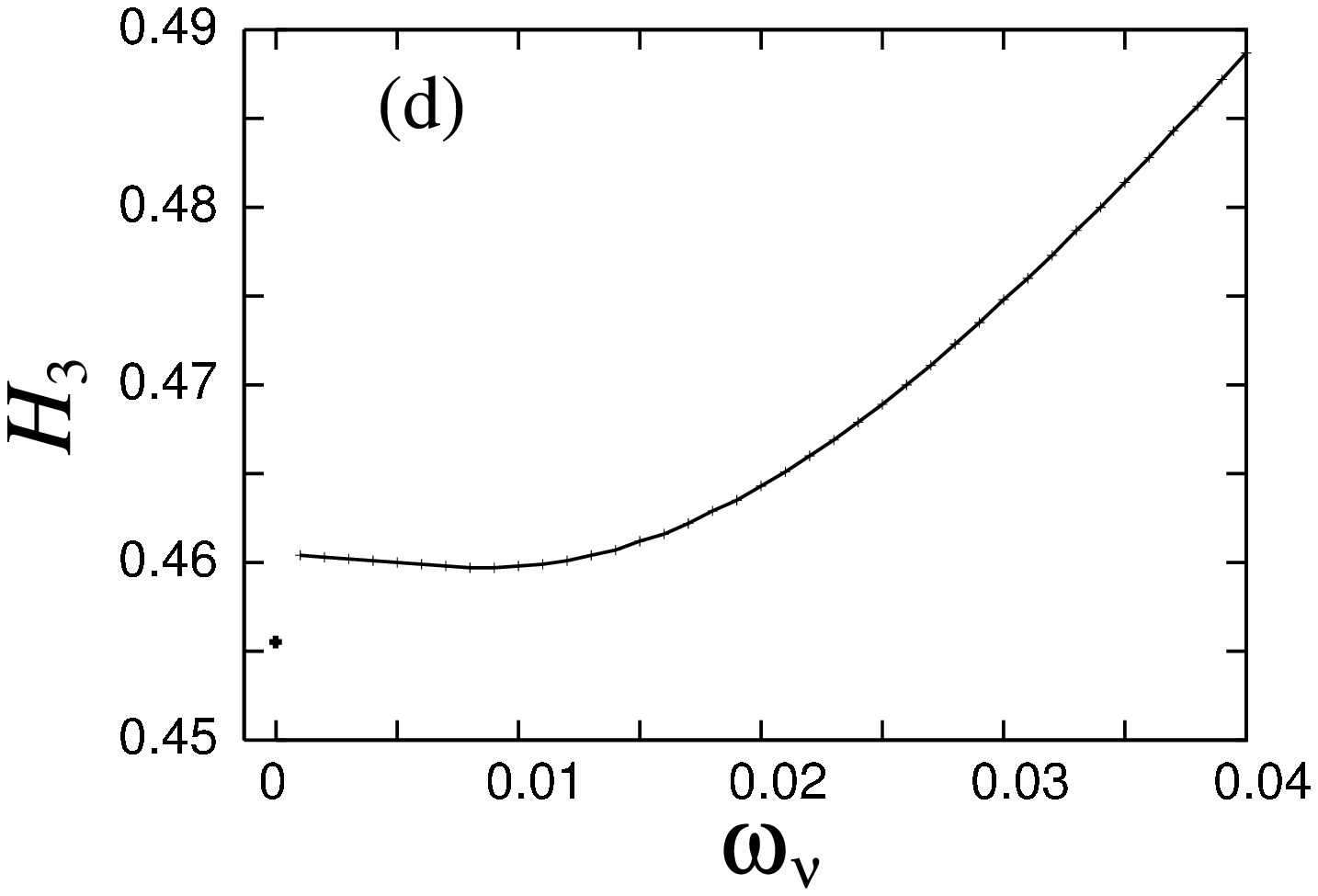} \\
& \\
\end{tabular}
\caption{Response of the four reduced CMB observables to the variation of
$\omega_{\nu}$. The isolated points show the values at $\omega_\nu=0$, 
which do not connect to the $\omega_\nu\ne 0$ values smoothly.}
\label{fig:on-observables}
\end{center}
\end{figure}

We draw the response of $O_i$ 
against the  variation of $\omega_{\nu}$ for the range
$\omega_{\nu}=0$ to 0.04 in Figures. 5. 
Note that an increase in
$\omega_{\nu}$ accompanies a decrease in $\Omega_{\Lambda}$
as we keep $\Omega_{\rm tot}=1$ and $\omega_m=\omega_{\rm cdm}+\omega_b$ 
fixed\footnote{Which variables are to be used is merely a matter 
of the convention. We chose the ones with which the effect of massive
neutrinos is more clearly visible.}.
We see small glitches from $\omega_{\nu}=0$ to the neighbouring point
in Figures. 5 (b), (c) and (d). 
This is probably a numerical
artefact caused by the implementation of the massive neutrino
subroutine in the CMBFAST code, and
we ignore these glitches since they are much smaller than the 
errors of the CMB data.

We observe that the response of the four observables against
$\omega_\nu$ changes at around  $\omega_\nu\approx 0.015$.
As $\omega_{\nu}$ increases beyond it, the decrease of $\ell_1$
becomes gentle; the $H_1$, which increases with $\omega_{\nu}$
up to $\omega_{\nu}=0.017$,
turns to decrease. 
$H_2$ and $H_3$ change little ($<0.5$\%)
between $\omega_{\nu}=0$ and 0.015, but
then begin to increase.
We can understand this turning point as the competition between
neutrino free streaming and recombination.
Neutrinos become non-relativistic before the
recombination when $\omega_{\nu} \gtrsim 0.017$ and they become
non-relativistic after the recombination when $\omega_{\nu} \lesssim
0.017$. We can show that the behaviours, at least for $\ell_1$ and
$H_1$, are quantitatively understood by simple analytical
considerations, but let us defer this problem to the next section.

Here, we are concerned with the problem how the constraint on the
neutrino mass is obtained from the CMB data alone,
given observational and empirical information of $\ell_1$,  $H_1$,
$H_2$ and $H_3$.
We argue that we cannot derive a constraint for $\omega_{\nu} <
0.017$ but an upper limit likely exists at some neutrino mass
in the region $\omega_{\nu} \gtrsim 0.02$.

We first consider $\omega_{\nu} < 0.017$. In this regime, as seen in 
Figure 5,
$\ell_1$ decreases and $H_1$
increases while $H_2$ and $H_3$ change little with increasing 
$\omega_\nu$.
The change induced by $\omega_\nu$ in  $\ell_1$
is significant, but according to Eq.~(\ref{eq:deriv_l1}) it
is cancelled to a large degree by a decrease of $h$ [Fig. 2 (c)], as seen in
Figure~3 (a). 
For $\omega_{\nu}=0.015$, say, we need $h$
to decrease from 0.69 to 0.58, but this change is harmless.
The decrease of $h$, however, causes an increase of $H_1$ (see 
(\ref{eq:deriv_H1}))
in addition to the direct increase due to $\omega_{\nu}$.  The increase of
$H_1$ is cancelled by decreases of $n_s$ and $\omega_b$,
whereas those two decreases tend to cancel in $H_2$ and $H_3$.
The error allowance of $H_1$ is large enough that a good
cancellation is not required, and hence it is easy to make the induced
changes of $H_2$ and $H_3$ cancelled to within their error allowances.
The large error of $H_1$ arises primarily from the
cosmic scatter, $\sigma\sim\sqrt{2/(2\ell+1)}$, 
in small $\ell$ modes (which we estimate to give
$\delta H_1\approx \pm 0.5$); so it seems unlikely that it will be reduced
greatly in data expected in the future. 
Therefore, unless external observational data are introduced
we cannot derive a
constraint on the neutrino mass density for $\omega_{\nu}$ substantially
smaller than  0.017, consistent with the flat $\chi^2$ dependence
around $\omega_\nu=0$ in Figure 1.
This will remain to be true even if the quality of the CMB data
is improved.

When $\omega_{\nu} > 0.017$, massive neutrinos contribute to
increase $H_2$ and $H_3$ 
as seen in Figures. 5
(c) and (d) in addition to
a further decrease of $\ell_1$.
Looking at Eq.~(\ref{eq:deriv_H2}) and Eq.~(\ref{eq:deriv_H3}),
the increase in $H_2$ and $H_3$
due to massive neutrinos may be compensated by either increasing
$\omega_b$ or decreasing $n_s$.
Actually, as shown in
Figure~2 
(a) and (e), the decrease
of $n_s$ occurs to minimise $\chi^2$. This is owing to a steeper
increase of $H_3$ than $H_2$ with the increase of $\omega_{\nu}$:
$\Delta  H_3/\Delta\omega_{\nu} > \Delta H_2/\Delta\omega_{\nu}$
in Figure. 5 (c) and (d).
Such increases are more
efficiently compensated by the decrease of $n_s$ than by the increase of
$\omega_b$, as read from 
Eqs.~(\ref{eq:deriv_H2}) and
(\ref{eq:deriv_H3}), which indicate that $\Delta H_3/\Delta n_s > \Delta
H_2/\Delta n_s$ whereas $|\Delta H_3/\Delta \omega_b| < |\Delta 
H_2/\Delta
\omega_b|$. [NB: $\omega_b(\omega_\nu=0.017)-\omega_b(\omega_\nu=0)$ 
is negative 
for the reason discussed in the above paragraph, but turns to increasing
 for $\omega_{\nu}>0.017$ to collaborate in the requirement.]
In other words, massive neutrinos enhance the multipoles
more on smaller scales (larger $\ell$), which causes an effect similar to
the increase of $n_s$ than the decrease of $\omega_b$, which increases
even peaks more strongly.

In passing, it is worth noting that $\omega_\nu$ and $n_s$ are negatively
correlated in this argument, in contrast to the naive expectation of
the positive correlation from the effect of massive neutrinos that diminish 
the small scale power. The latter implies that the limit on the neutrino mass
loosens for increasing $n_s$ (e.g., \cite{Fukugita:2000}). 
The CMB argument works in the opposite way.

The cancellation of the effect due to $\omega_\nu$
in the acoustic peaks by decreasing $n_s$ increases the
large-scale amplitude significantly, as
is manifest in a large coefficient of $\Delta n_s/n_s$ in
Eq.~(\ref{eq:deriv_H1}). With a tight error allowance for $H_2$
the decrease of $n_s$ compels $H_1$ to decrease largely, as seen in
Figure 3 (b),  
and to push down $H_1$ below the allowed error range
($H_1\ge 6.2$ at 1.5 $\sigma$)  at
around $\omega_{\nu} \sim 0.02$, while 
$H_2$ and $H_3$ stay within the
boundary of errors given in Eqs.~(\ref{eq:error2_H2}) 
and (\ref{eq:error2_H3}).
This corresponds to
the upper limit of $\omega_{\nu}$ we obtained, i.e.,
$\omega_{\nu} < 0.021$ (at 95\%), 
in a numerical study of the $\chi^2$ test.

Let us visit briefly the possibility of varying $\tau$ to
increase $H_1$. From Eq.~(\ref{eq:deriv_H1}),
a large decrease of $\tau$ would
make it increase without disrupting $\ell_1$, $H_2$ and $H_3$. However, $\tau$ 
can
not be reduced as much as one wants, as displayed in
Figure~2 (d). 
The observed high
amplitude at the lowest multipoles of the TE mode needs a non-negligible
amount of the reionization optical depth.

We may also ask whether the inclusion of the tensor perturbations change
the limit. Hu et al. \cite{Hu:2000ti} give
\begin{equation}
\Delta H_1\approx -5\frac{r_t}{1+0.76r_t}
\end{equation}
where $r_t=1.4[\Delta T_{10}^{(T)}/\Delta T_{10}^{(S)}]^2$ is the tensor
to scalar ratio at $\ell=10$. This means that the inclusion of the
tensor mode collaborates to lower $H_1$, and thus only tightens the limit
on the neutrino mass density.

These considerations show that one can derive the limit on the
neutrino mass density of the order of $\omega_{\nu} \sim 0.02$
   from WMAP data alone. They also show that the limit may be
improved to $\omega_{\nu} \sim 0.017$ with the use of improved
CMB data, but it would not be easy to go beyond.
Even with the extremely high precision data anticipated from PLANCK,
the limit we expect will be  $\omega_{\nu} < 0.013$ at the
95\% confidence level\footnote{In this estimate we use the assumed
CMB data that lie around the best $\Lambda$CDM solution 
for the vanishing neutrino mass 
with the error being the cosmic variance. We used our data base
to search for the $\chi^2$ minimum.}: the increase of $\chi^2$ is very slow
for $\omega_\nu\lesssim 0.01$\footnote{Kaplinghat 
et al.\cite{Kaplinghat:2003bh} proposed to use the
deflection angle power spectrum from weak gravitational lensing to
give a stronger constraint on $m_\nu$. We do not take this into
accout in the present consideration.}.

The efficient way to improve the limit on   $\omega_{\nu}$
is to introduce observations that constrain the Hubble constant,
either directly or indirectly, from below. This is because
the most prominent effect caused by light neutrino is to
change the position of the first peak and it is absorbed into
a lowering shift of the Hubble constant.
Should one require
that $H_0>65$ km s$^{-1}$Mpc$^{-1}$, a significantly stronger limit
of the order of $\omega_{\nu} \lesssim 0.01$ 
would be derived even with the
current CMB data\footnote{With this lower limit on $H_0$, the 
global $\chi^2$ minimum is given by the unphysical solution
that gives unreasonably large reionisation optical depth.
Our statement in the text excludes this possibility.}.

\section{Analytic considerations on the effect of massive neutrinos}
\label{sec:analytic}

\subsection{The position of the first peak}
\label{sec:scales}
Here, we attempt to understand the effect of massive neutrinos on the
reduced CMB observables.
We may take the epoch when the neutrino of mass $m_{\nu}$
becomes nonrelativistic as its momentum $p_\nu\sim m_\nu$, i.e.,
$T_{\nu,\rm nr} = m_{\nu}/3$. The  corresponding redshift is
\begin{eqnarray}
1+z_{\rm nr} &=& \frac{T_{\nu,\rm nr}}{T_{\nu,0}} \\
&=& 1.99 \times 10^3 (m_{\nu}/{\rm eV}) \\
&=& 6.24 \times 10^4 \omega_{\nu}, \label{eq:z_nr}
\end{eqnarray}
where 
$\sum m_{\nu} = 3 m_{\nu}$ is used for the last equality.
This is compared with the redshift
at recombination $z_{\rm rec}=1088$ \cite{Spergel:2003cb}, which
is insensitive to the mass of neutrinos.
Neutrinos become non-relativistic before
recombination, i.e., $z_{\rm nr} > z_{\rm rec}$, if
\begin{eqnarray}
\omega_{\nu}  \gtrsim 0.017,
\end{eqnarray}
but otherwise they remain relativistic and freely stream till 
post recombination epochs.
This $\omega_{\nu}$ corresponds approximately to the turning points
of the curves of $\ell_1$, $H_1$, $H_2$ and $H_3$
observed in Figure 5. 

We denote the energy density in the form
$\omega\equiv\Omega h^2= \rho h^2/\rho_{\rm cr,0}$,
where the critical density
$\rho_{\rm cr,0}=3M_{\rm pl} H^2_0$ 
with the Planck mass defined by
the gravity scale $M_{\rm pl}^2=1/8\pi G$,
and the subscript $0$ expresses values at the present epoch.
The matter and photon energy densities are 
\begin{eqnarray}
\rho_m(a) h^2/\rho_{\rm cr,0} = \omega_{m,0} \left( \frac{a}{a_0}
\right)^{-3}, \hskip5mm
\rho_{\gamma}(a) h^2/\rho_{\rm cr,0} = \omega_{\gamma,0} \left(
\frac{a}{a_0} \right)^{-4},
\end{eqnarray}
where the present photon energy density $\omega_{\gamma,0} = 2.48
\times 10^{-5}$ for
$T_{\gamma,0} = 2.725$ K \cite{Mather:1998gm}.
The neutrino energy density is
\begin{eqnarray}
\rho_{\nu}(a) h^2/\rho_{\rm cr,0} &=& \frac{45}{\pi^4} \left(
\frac{4}{11}
\right)^{4/3}\omega_{\gamma,0}\left(\frac{a}{a_0}\right)^{-4}
\int_0^{\infty} \sqrt{x^2 + y^2} x^2 (e^x +1)^{-1}dx, \label{eq:rhonu}
\end{eqnarray}
where
\begin{equation}
y=m_{\nu}(11/4)^{1/3}(a/a_0)T_{\gamma,0}^{-1} ,
\label{eq:y}
\end{equation}
and $x$ is the
normalised momentum variable and three flavours of neutrinos
are assumed to have a degenerate mass.
The vacuum energy is
\begin{eqnarray}
\rho_{\Lambda}(a) h^2/\rho_{\rm cr,0} &=& \omega_{\Lambda,0}   \\
&=& h^2 - \omega_{m,0} -\omega_{\nu,0},
\end{eqnarray}
for the flat universe.
The total energy density is $\rho_{\rm tot} = \rho_m + \rho_{\gamma} +
\rho_{\nu} + \rho_{\Lambda}$.
With $\rho_{\rm tot}$,
the cosmic expansion rate $H=\dot a/a$ is given by
$H^2 = \rho_{\rm tot}/3 M_{\rm pl}^2$, which is used to evaluate the
conformal time $\eta$,
\begin{eqnarray}
\eta(a) = \int \frac{dt}{a} = \int_0^a \frac{da^{\prime}}{a^{\prime 2}
H}\ .
\end{eqnarray}

The position of the $m$-th peak $\ell_m$ is determined from
that of the acoustic peak $\ell_A$ and the phase shift $\phi_m$,
which depends weakly on $m$ \cite{Hu:2000ti},
\begin{eqnarray}
\ell_m = \ell_A(m-\phi_m),
\label{eq:ell-m}
\end{eqnarray}
where the acoustic scale is defined by 
\begin{eqnarray}
\ell_A = \pi\frac{r_{\theta}(\eta_{\rm rec})}{r_s(\eta_{\rm rec})},
\end{eqnarray}
with $r_s(\eta_{\rm rec})$ the sound horizon at the recombination epoch
and $r_{\theta}(\eta_{\rm rec})$ is the comoving
angular diameter distance to the last scattering surface,
$r_{\theta}(\eta_{\rm rec})=\eta_0-\eta_{\rm rec}$ in the flat universe.
The sound horizon is given by
\begin{eqnarray}
r_{s}(a) \equiv \int_0^{\eta(a)} c_s d\eta = \int_0^a c_s(a^{\prime})
\frac{da^{\prime}}{a^{\prime 2} H},
\end{eqnarray}
where the sound speed $c_s^2=(1/3)(1+R)^{-1}$ with
$R=3\rho_b/4\rho_{\gamma}=3a\omega_{b,0}/4\omega_{\gamma,0}$. The
$c_s$ depends only on photons and baryons, and the effect of neutrino
masses enters into the sound horizon only through the modification of the
expansion law. The phase shift in Eq.~(\ref{eq:ell-m}) 
arises from the decay of gravitational potential
due to radiation growth suppression when the universe
is not fully matter dominated, which 
later modifies the gravitational redshift that the photons would
otherwise suffer from the Sachs-Wolfe effect \cite{Hu:1994jd}. 
This is sometimes called
the early integrated Sachs-Wolfe effect.
 The evaluation of the integral gives
$\ell_A \sim 300$, which is considerably larger than the physical 
position
of $\ell_1$: the difference is ascribed to
the phase shift $\phi_1$, which is estimated in what follows.

The enhancement of the amplitude for scales between the first
acoustic peak and the horizon crossing at the matter domination
due to the early integrated Sachs-Wolfe effect makes the first peak formed at
a scale larger than the acoustic peak.
An accurate evaluation of the phase shift $\phi$
requires the full solution of the coupled Boltzmann equations.
Instead, we use the fitting formula given in Ref.~\cite{Hu:2000ti},
\begin{eqnarray}
\phi_1 \approx 0.267 \left( \frac{r_{\rm rec}}{0.3} \right)^{0.1},
\label{eq:phi}
\end{eqnarray}
where $r_{\rm rec}$ is the radiation-to-matter energy ratio $r \equiv
\rho_r/\rho_m$ at the recombination. The appearance of the
radiation-to-matter energy ratio as the prime variable is
motivated by the physics of the integrated Sachs-Wolfe effect
\cite{Hu:1994jd}. Precisely speaking, this fitting formula is
given for massless neutrinos, but it is expected to be valid
for massive case provided that $r_{\rm rec}$ is modified appropriately,
because the effect of massive neutrinos on the integrated Sachs-Wolfe
effect is primarily through the change of $r_{\rm rec}$.
A larger radiation-to-matter energy ratio leads to
a larger enhancement and hence a larger phase shift as indicated by
Eq.~(\ref{eq:phi}). Massive neutrinos with $\omega_\nu\gtrsim 0.017$
act in a way to suppress this 
effect.

The ratio $r_{\rm rec}$ in the presence of massive neutrinos is calculated
as follows.
We take neutrinos that have momentum larger than $m_{\nu}$ as
radiation, and those having smaller momentum as
matter. Accordingly, we split $\rho_{\nu}$ into the radiation
component $\rho_{\nu,r}$ and the matter component $\rho_{\nu,m}$, as
\begin{eqnarray}
\rho_{\nu,r}(a) h^2/\rho_{\rm cr,0} &=& \frac{45}{\pi^4} \left(
\frac{4}{11}
\right)^{4/3}\omega_{\gamma,0}\left(\frac{a}{a_0}\right)^{-4}
\int_y^{\infty} \sqrt{x^2 + y^2} x^2 (e^x +1)^{-1}dx,
\label{eq:rhonu_r} \\
\rho_{\nu,m}(a) h^2/\rho_{\rm cr,0} &=& \frac{45}{\pi^4} \left(
\frac{4}{11}
\right)^{4/3}\omega_{\gamma,0}\left(\frac{a}{a_0}\right)^{-4}
\int_0^{y} \sqrt{x^2 + y^2} x^2 (e^x +1)^{-1}dx, \label{eq:rhonu_m}
\end{eqnarray}
by dividing the integration range at the value in Eq.~(\ref{eq:y}).
The radiation-to-matter energy ratio is calculated as
\begin{equation}
\xi=(\rho_{\gamma}+\rho_{\nu,r})/(\rho_m+\rho_{\nu,m}),
\label{eq:r}
\end{equation}
  which is
used to compute $\phi_1$ in Eq.~(\ref{eq:phi}).

\begin{figure}
\includegraphics{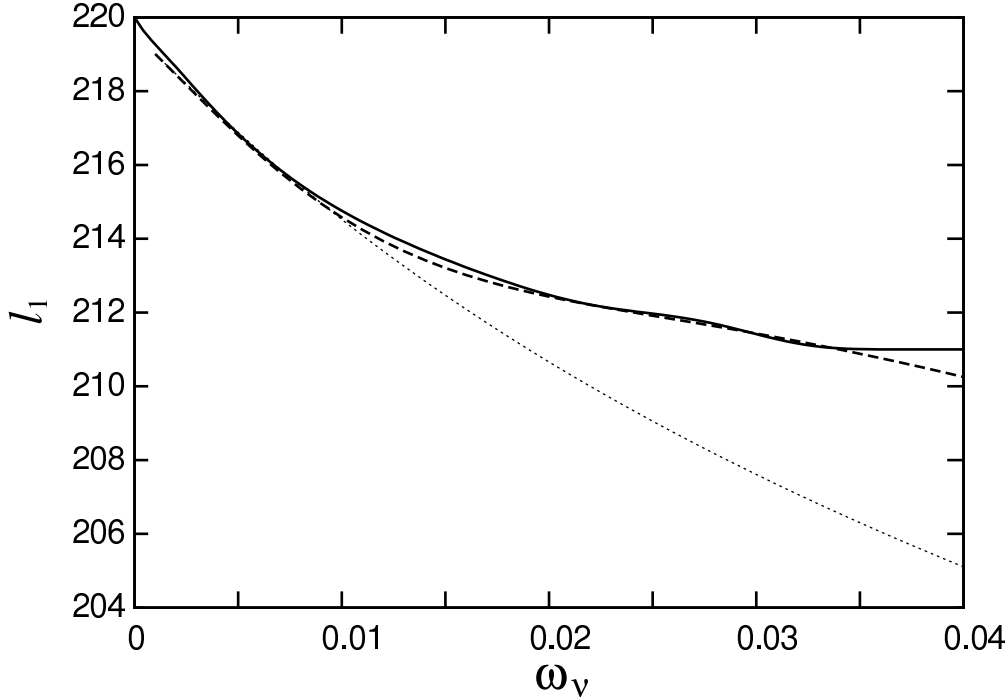}
\caption{Dependence of $\ell_1$ on $\omega_{\nu}$ 
calculated from Eq.~(\ref{eq:ell-m}) (dashed line),
as compared with the accurate numerical solution (solid line).
The dotted line shows the case when the effect of massive neutrinos
on the early Sachs-Wolfe effect is ignored.}
\label{fig:l_Aandl_1}
\end{figure}

The first peak position thus calculated as a function of $\omega_{\nu}$
is shown in Figure~\ref{fig:l_Aandl_1} together with the curve from the
full numerical computation presented earlier.
The agreement is excellent, validating the prescription described here.
For a reference we also draw the case where the phase shift
is fixed at the zero-neutrino-mass value,
$(1-\phi_1) \sim 220/300$.
This curve agrees with the accurate result for small neutrino
masses, but starts deviating from
$\omega_{\nu} \approx 0.015$, i.e., when neutrinos become
nonrelativistic before the recombination epoch.
This stands for the error that we count neutrinos as
radiation even when they are
non-relativistic at the recombination, and hence, overestimates the 
early
integrated Sachs-Wolfe effect, so does the phase shift $\phi_1$.
This consideration demonstrates that the change of the slope
in $\ell_1$ at $\omega_{\nu}\approx 0.017$
is a result of the reduction of the early integrated
Sachs-Wolfe effect by the neutrinos that become nonrelativistic
before the recombination epoch.

\subsection{Hights of the acoustic peaks}
\label{sec:peaks}

\begin{figure}
\includegraphics{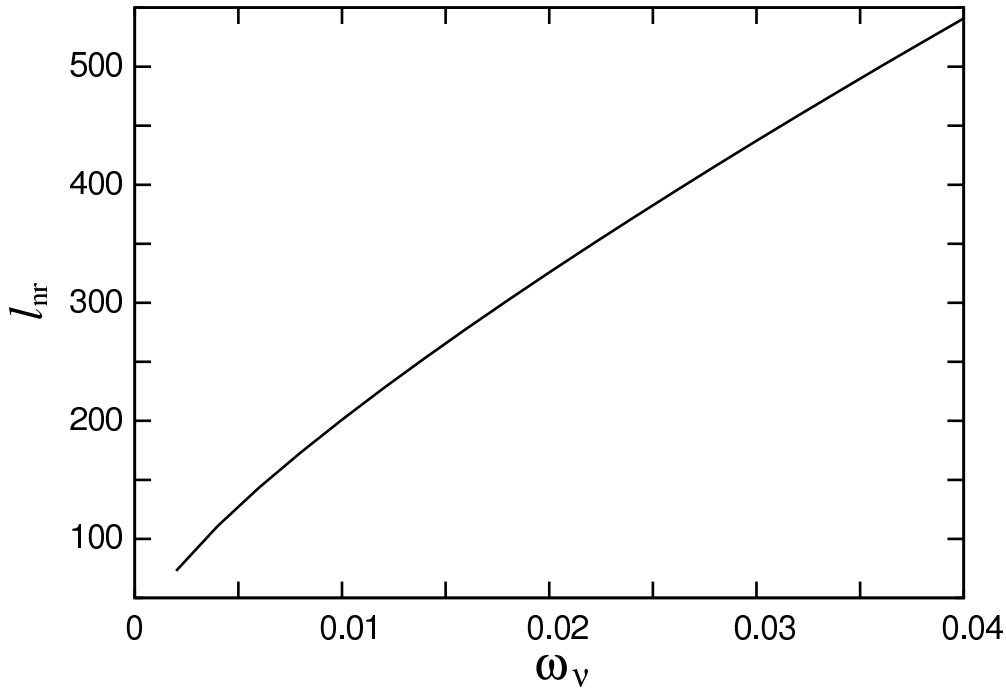}
\caption{Multipoles corresponding to the neutrino free-streaming
scale.}
\label{fig:nrscale}
\end{figure}

It is known that free-streaming of massive neutrinos causes a 
larger decay in the gravitational potential $\Phi$. This drives
the acoustic oscillation of the baryon-photon fluid more
strongly, so that  
the amplitude of temperature anisotropies within  
the free-streaming scale is enhanced through the
monopole term $\Theta_0+\Psi$ in the harmonic expansion of the temperature
perturbations \cite{Dodelson:1995es}.
The conformal time corresponding to the free-streaming scale
is calculated as  $\eta_{\rm nr} = \eta(a_{\rm nr})$ where $a_{\rm nr}$ is known from
Eq.~(\ref{eq:z_nr}). This is the
distance over which relativistic neutrinos move freely.
The multipole $\ell_{\rm nr}$ corresponding to this scale 
is~\cite{Hu:1994jd}:
\begin{eqnarray}
\ell_{\rm nr}&\simeq&\frac{2\pi r_{\theta}(\eta_{\rm rec}) }{\eta_{\rm nr}},
\end{eqnarray}
which we show in Figure~\ref{fig:nrscale}
for $\omega_{m,0}=0.14$ and
$h=0.69$, and $z_{\rm rec}=1088$. 
The multipole amplitudes for $\ell>\ell_{\rm nr}$ are affected by free 
streaming.
For $\omega_{\nu} > 0.017$, the amplitude
on the scale $\ell > 300$ is enhanced \cite{Dodelson:1995es}.
This means that only the second and higher peaks receive the effect.

The first peak receives little the effect of the
decay of gravitational potential, and
the variation of $H_1$ with $\omega_\nu$ is understood by a simple
consideration.
The gentle increase of $H_1$ for
$\omega_{\nu} \lesssim 0.017$ is understood by the decrease of
$\Omega_{\Lambda}$ to compensate the neutrino energy density
in the flat universe 
and an associated decrease of the integrated Sachs-Wolfe effect
from the late domination of $\Lambda$,
which enhances $C_{10}$. The effect continues to the region
$\omega_{\nu} \gtrsim 0.017$, but in this regime massive neutrinos
act as the nonrelativistic dark matter at recombination
and the effect from the increase of the amount of matter overcomes;
hence $H_1$ begins to decrease
[$\Delta H_1/\Delta \omega_m<0$ as seen in Eq.~(\ref{eq:deriv_H1})].
This indicates that
$\omega_{\nu} \sim 0.017$ is again the turning point, as we saw
in Figure 5. 
In what follows we verify this reasoning by a more quantitative
argument.

Our strategy is to reduce the theory with massive neutrinos to an effective, 
mock
theory without massive neutrinos, for which we have an established
understanding \cite{Hu:1994jd,Hu:2000ti}.
If neutrinos are light they are taken as radiation,
and if heavy, they are regarded  as matter.
For $\omega_{\nu}\approx 0.017$,
they contribute as both matter and radiation, and are handled
by splitting the neutrino energy density into the radiation and matter
parts as in Eqs.~(\ref{eq:rhonu_r}) and (\ref{eq:rhonu_m}).
We count the matter part of neutrinos at the recombination as
additional ``CDM''. We then have the effective matter density,
\begin{eqnarray}
{\tilde\omega}_{m} = \omega_{m,0} +
\frac{\rho_{\nu,m}(a_{\rm rec})}{\rho_{\nu,r}(a_{\rm rec})+\rho_{\nu,m}(a_{\rm rec})
} \omega_{\nu,0}, \label{eq:om_eff}
\end{eqnarray}
where $a_{\rm rec}=1/1089$; see Figure 8 (a). 

In order to mimic the true matter-radiation equality epoch
in the theory without having massive neutrinos, we
try to vary the effective number of neutrino species 
$N_{\nu}$. This ensures nearly the
same amount of the early integrated Sachs-Wolfe effect
generated in the massless neutrino world.
The scale factor at
the equality $a_{\rm eq}$ as a function of $\omega_{\nu}$ is calculated from
the condition $\xi(a_{\rm eq})=1$ where $\xi$ is
defined by Eq.~(\ref{eq:r}). The result is shown in Figure 8 (b)
\footnote{A gentle increase for small $\omega_{\nu}$ in
Figure 8 (b) is caused by the increase in the radiation
component of the neutrino energy density $\rho_{\nu,r}$ relative to the
matter counterpart $\rho_{\nu,m}$ for small $\omega_{\nu}$. Note that
$\rho_{\nu,r}$, defined by Eq.~(\ref{eq:rhonu_r}), is not necessarily
monotonically decreasing as a function of neutrino mass.}.
 From the conventional calculation giving $1+z_{\rm eq} = a_{\rm eq}^{-1} = 80950
\omega_m/(2+0.454 N_{\nu})$ for $\omega_{\nu}=0$, the effective
$N_{\nu}$ we want is
\begin{eqnarray}
N_{\nu} = \frac{80950 {\tilde\omega}_m a_{\rm eq}(\omega_{\nu,0}) - 
2}{0.454},
\label{eq:Nnu_eff}
\end{eqnarray}
which is shown in Figure 8 (c).

We also want to adjust $\Omega_{\Lambda}$ so that the integrated
Sachs-Wolfe effect in the $\Lambda$ dominated epoch
would be the same in the two universes.
Noting that the CMB perturbation depends on $h$ in the form $\Omega 
h^2$,
this may be accomplished by
shifting $h$. Because the flat universe requires
$(\omega_m+\omega_{\nu,0})h^{-2}+\Omega_{\Lambda} = 1$ for the
massive neutrino case, and $\omega_m \tilde h^{-2} + \Omega_{\Lambda} = 1$ for
the massless case, $h$ has to be reduced as
\begin{eqnarray}
\tilde h = h \sqrt{ \frac{\omega_m}{\omega_m + \omega_{\nu,0}} }.
\label{eq:h_eff}
\end{eqnarray}

We consider that the massless neutrino theory with these parameter 
shifts
captures the main features of the theory with massive neutrinos,
at least for the first acoustic peak.
In fact, as shown in Figure~\ref{fig:on-H1_th},
this mock theory reproduces  very well
the full calculation of $H_1$ with massive neutrinos.
In the same figure, we also show the two curves
calculated by adjusting either $\omega_m$ and $N_{\nu}$ alone or $h$ 
alone,
which represents, respectively, the effect of massive neutrinos as
matter or the increase of the vacuum energy.
The former curve is flat for $\omega_{\nu} \lesssim 0.017$ and
turns down henceforth.
These component curves demonstrate how $H_1$ is built.

\begin{figure}
\begin{center}
\begin{tabular}{cc}	
\includegraphics[width=7.5cm]{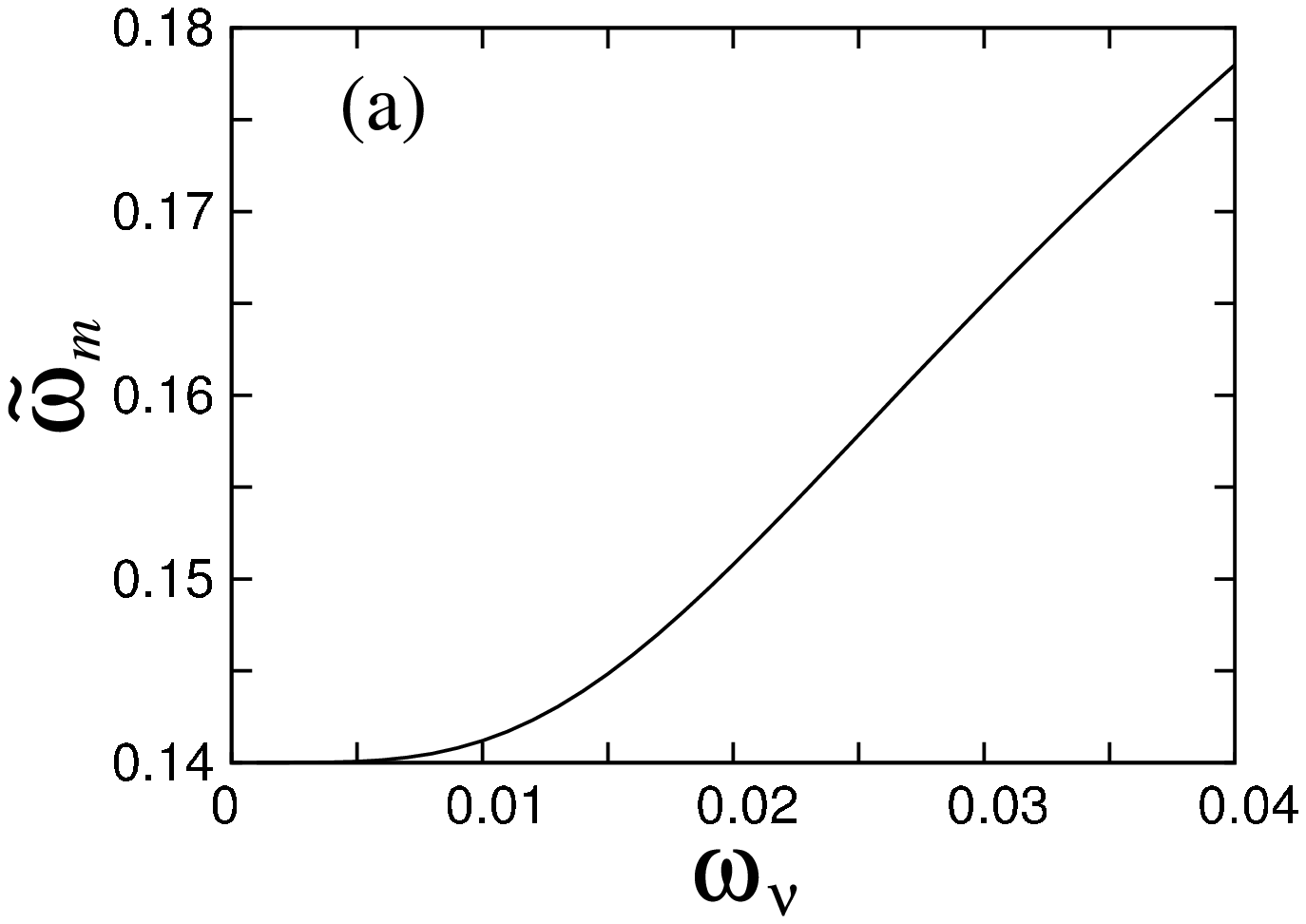} &
\hspace{0.5cm} \includegraphics[width=7.5cm]{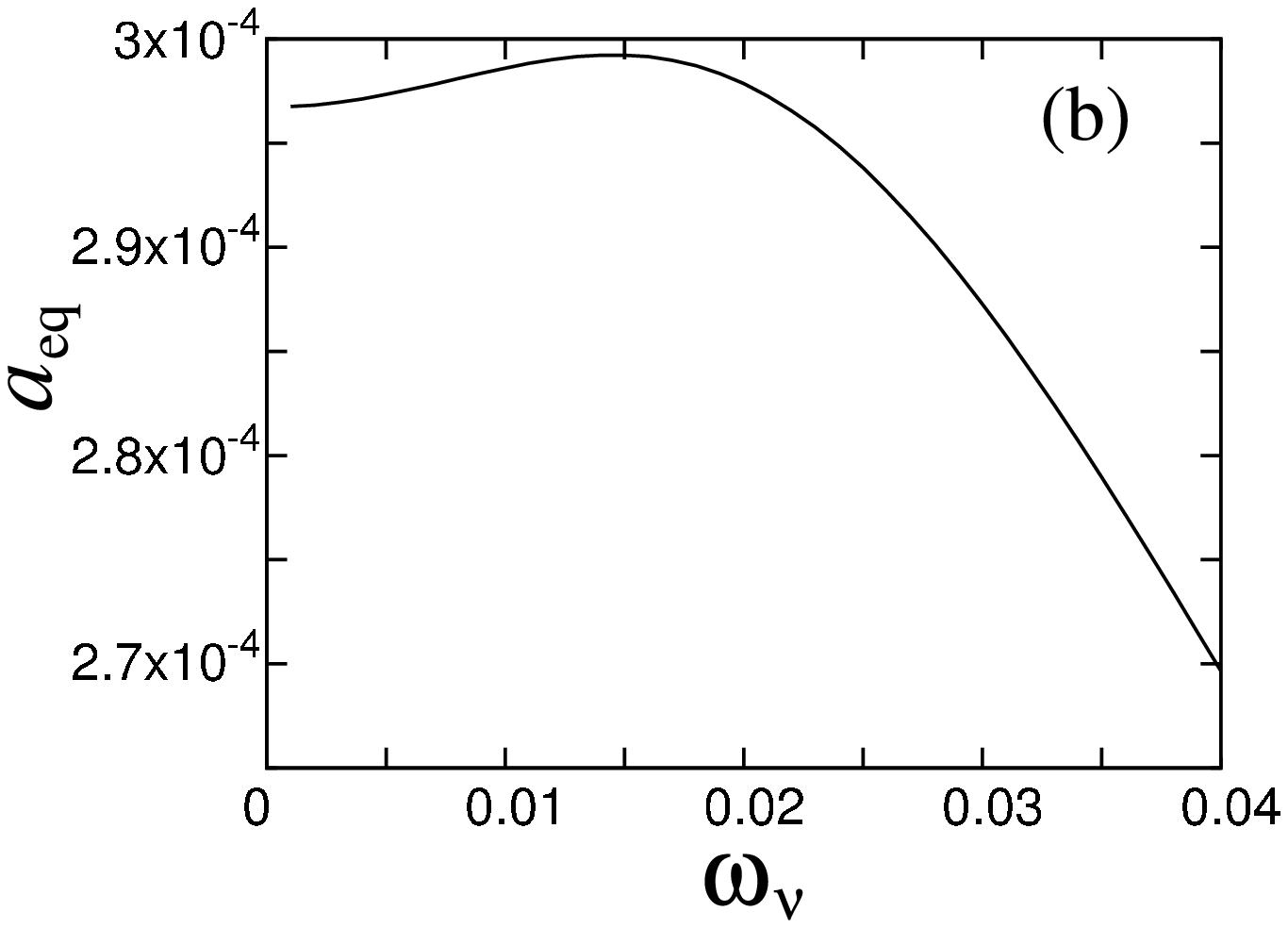} \\
& \\
\includegraphics[width=7.5cm]{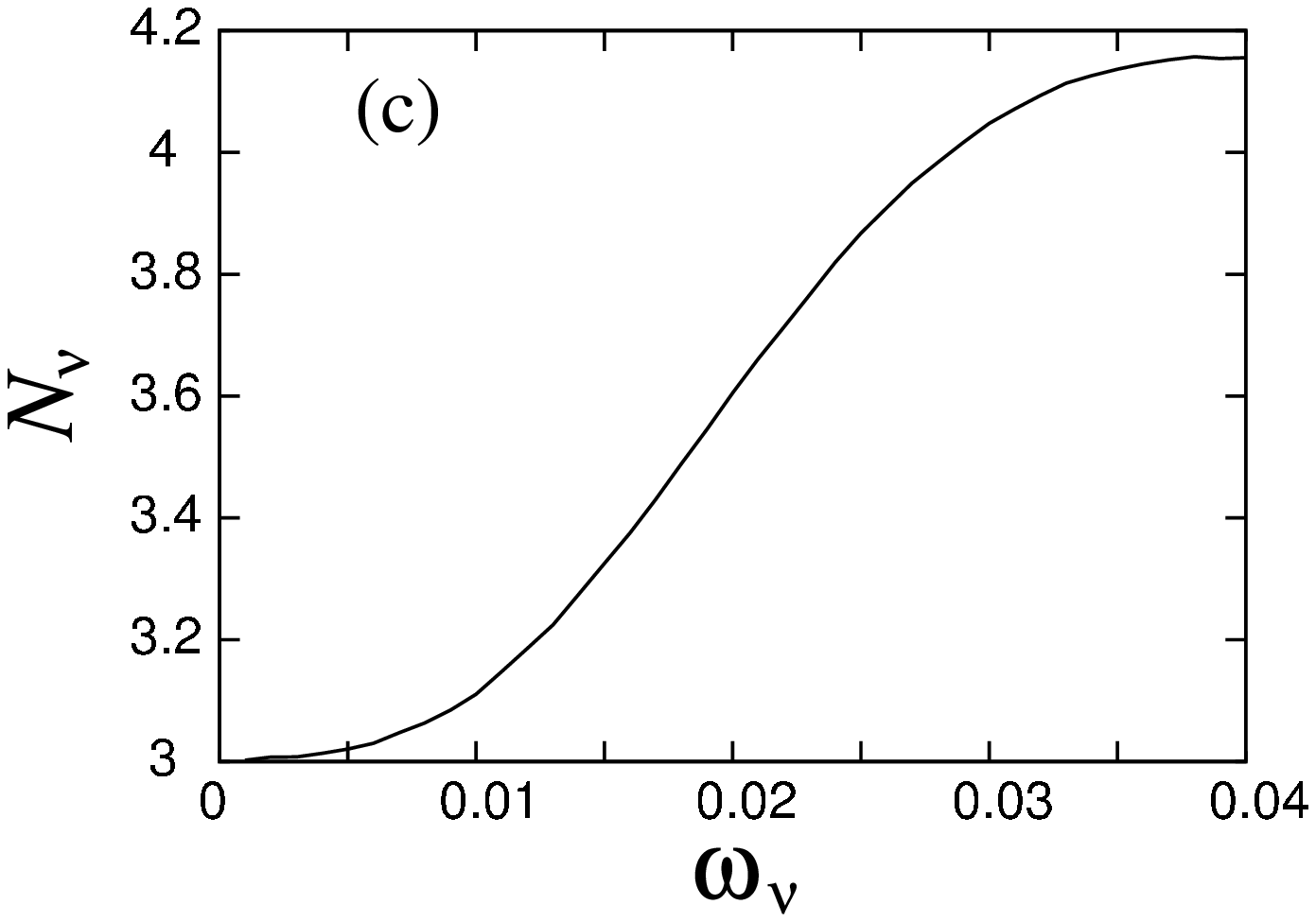} &
\hspace{0.5cm} \includegraphics[width=7.5cm]{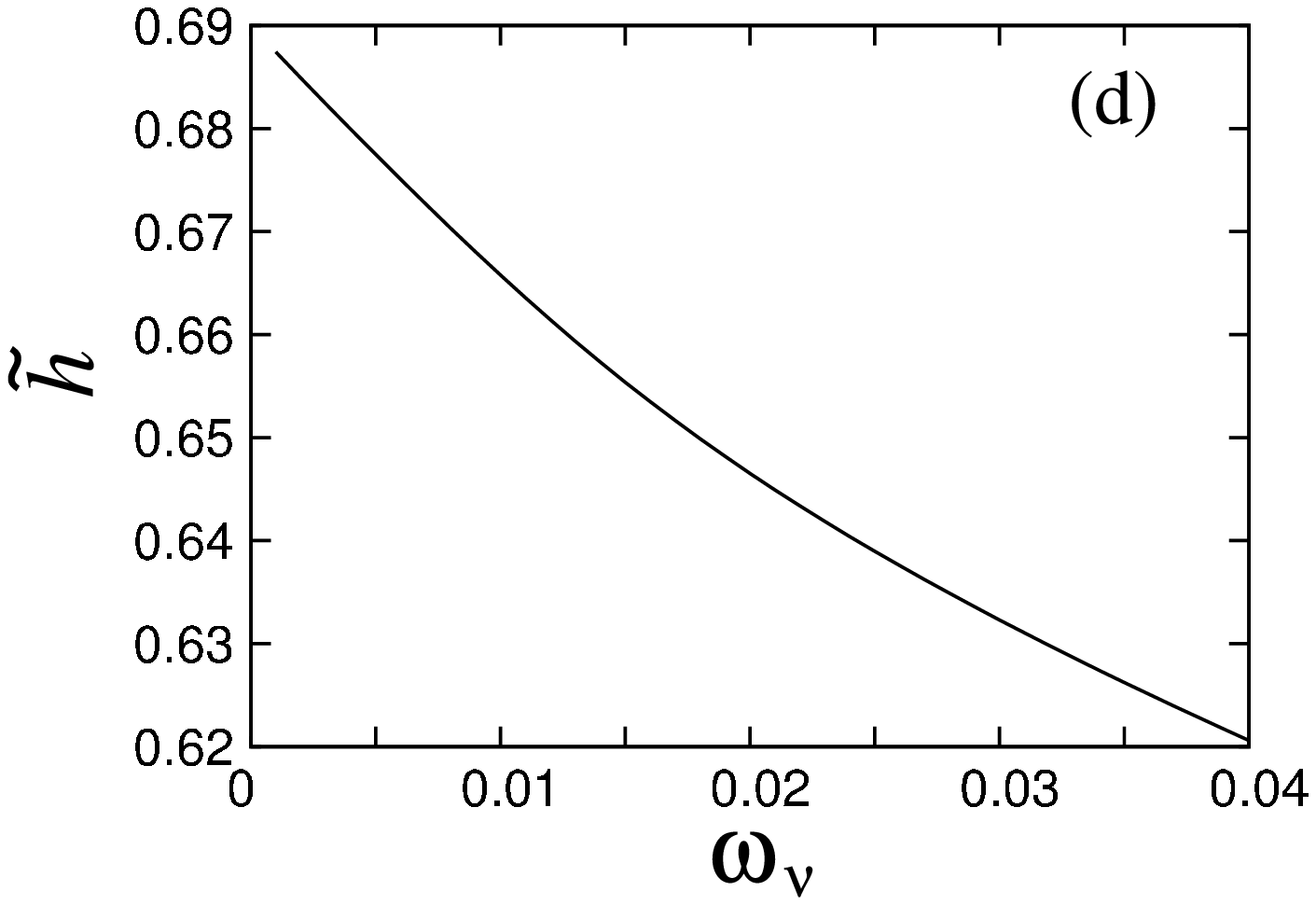} \\
& \\
\end{tabular}
\caption{Effective parameters of the massless neutrino theory
that are required to mock the massless neutrino 
world.}
\label{fig:parameters_eff}
\end{center}
\end{figure}

\begin{figure}
\includegraphics{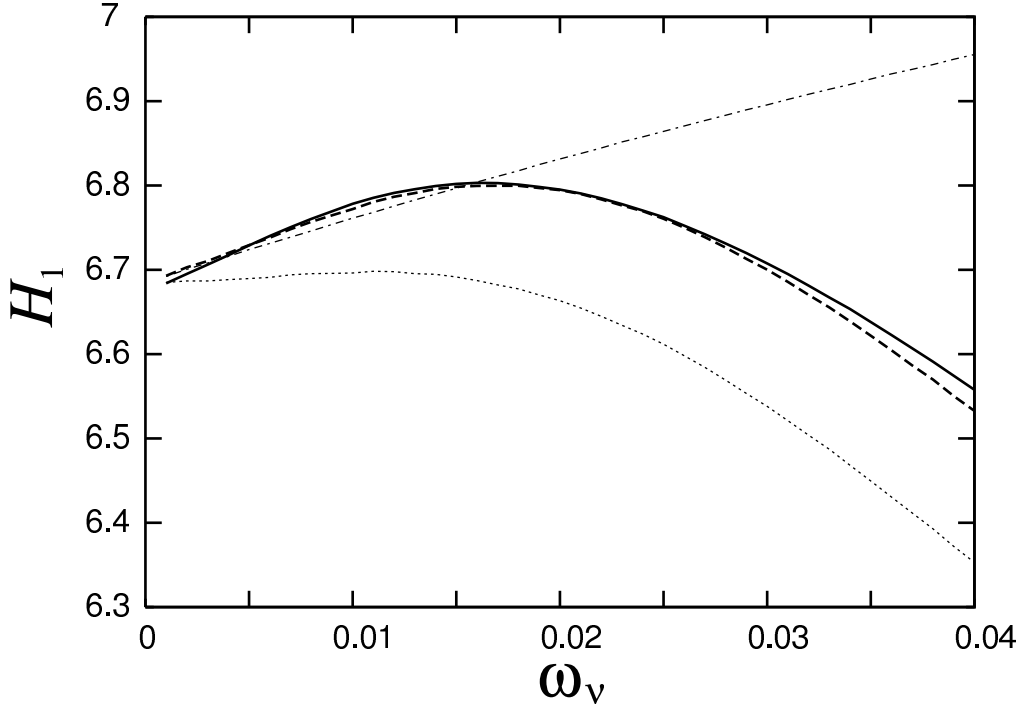}
\caption{Dependence of $H_1$ on $\omega_{\nu}$ predicted in the
mock massless neutrino theory (dashed line), as compared with
the true theory with massive neutrinos (solid line).
For illustration, the results with the theories, where only
the early integrated Sachs-Wolfe effect is mocked by changing $\omega_m$
and $N_{\nu}$ (dotted line) and only the late Sachs-Wolfe effect is 
mocked by changing  $h$, are also shown (dot-dashed line).  }
\label{fig:on-H1_th}
\end{figure}

\begin{figure}
\begin{center}
\begin{tabular}{cc}
\includegraphics[width=7.5cm]{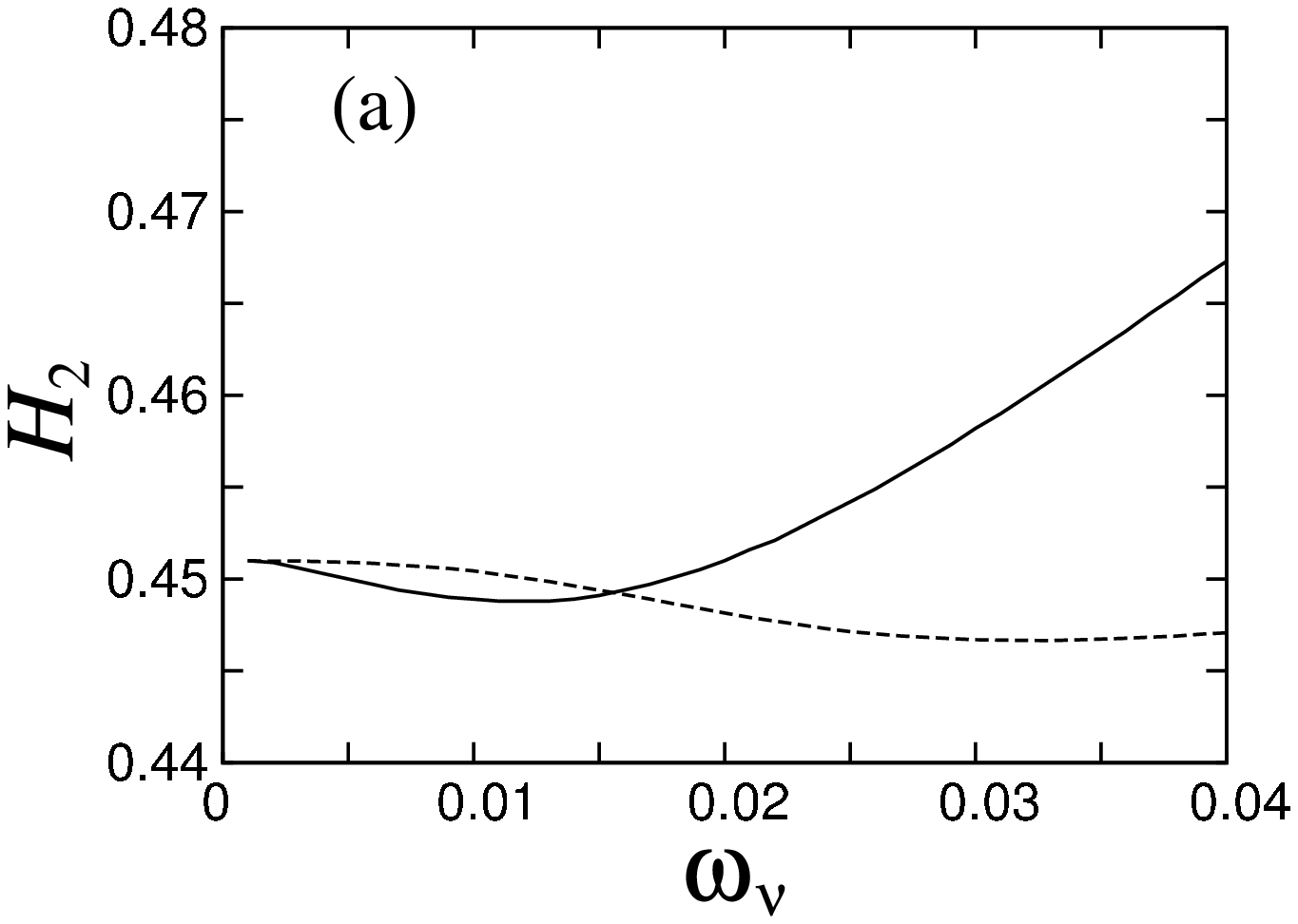} &
\hspace{0.5cm} \includegraphics[width=7.5cm]{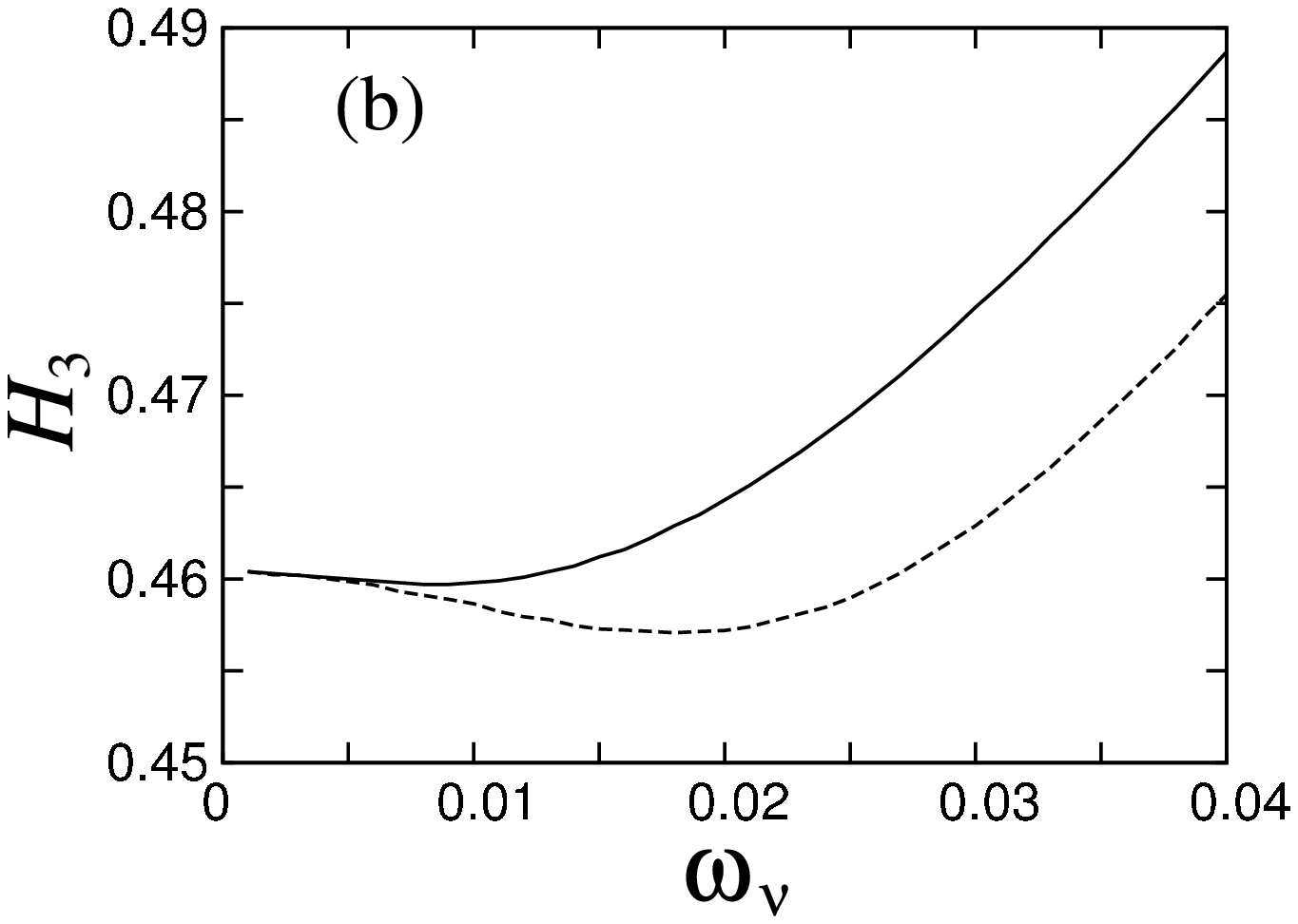} \\
& \\
\end{tabular}
\caption{Dependence of (a) $H_2$ and (b) $H_3$ on $\omega_{\nu}$ 
predicted in the
mock massless neutrino theory (dashed line), as compared with
the true theory with massive neutrinos (solid line).}
\label{fig:on-H2_th}
\end{center}
\end{figure}


The second and third peaks are enhanced by free streaming of massive
neutrinos~\cite{Dodelson:1995es}.
Ignoring this effect, however, we plot $H_2$ and
$H_3$ in Figure 10    
for the mock theory we used to reproduce $H_1$. Obviously, they do
not give the correct dependence for massive neutrinos, 
underestimating the true values of the changes in $H_2$ and
$H_3$ for $\omega_{\nu} \gtrsim 0.017$.
The effect of the potential decay
is more prominent in $H_2$ (Figure~\ref{fig:on-H2_th} (a))
to which the contribution of CDM is small but the baryon is the
major contributor (see Eq.~\ref{eq:deriv_H2}).
The increase of $H_3$ 
is partly accounted for by the modification of the  $\omega_m$
term that enters into $H_3$ in Eq.~(\ref{eq:deriv_H3}).
Dodelson et al. \cite{Dodelson:1995es} showed that the increase of 
the second and third peaks is understood by the potential decay.
We do not pursue our analysis further, as it would not give us 
more insight than that given by Dodelson et al.'s analysis.

\section{The neutrino mass constraint in non-flat universes}
\label{sec:nonflat}

We remove the assumption of $\Omega_{\rm tot}=1$
and study the constraint on the neutrino
mass in positive and negative curvature universes.
We made a $\chi^2$ minimum search only for a few values of 
$\omega_{\nu}$
close to the upper limit obtained in the flat universe, since the 
search is
time-consuming but an upper limit comparable to the one for
the flat universe is anticipated from an analytic argument. 
We only consider the
universe with $\Omega_{\rm tot}=1.02$, 1.04 (positive curvature)
and $\Omega_{\rm tot}=0.98$ (negative curvature), which are 
still allowed from WMAP.
The solutions that give a $\chi^2$ minimum are given for
$\omega_{\nu}=0$, 0.02, 0.025 and 0.03 in Table \ref{tab:chi2min_closed}
  for the positive curvature case ($\Omega_{\rm tot}=1.02$) and in Table
  \ref{tab:chi2min_open} for the negative curvature case.
The minimum $\chi^2$ is plotted in Figure~\ref{fig:chi2min}
presented earlier.
The figure shows that the $\chi^2_{\rm min}$ are slightly smaller for
the positive curvature and larger for the negative curvature
for a given $\omega_\nu(\ne 0)$.
We find, however, that this does not change the limit on the neutrino 
mass. For $\omega_\nu=0$, the universe of a slightly positive curvature
is somewhat more favoured, viz. 
$\chi^2({\rm flat},\omega_\nu=0)-\chi^2(\Omega_{\rm tot}=1.02,\omega_\nu=0)=1$,
as already known in earlier analyses \cite{Spergel:2003cb, Tegmark:2003ud}.
This decrease of $\chi^2$ at the global minimum compensates  
the decrease of $\chi^2$ seen at 
$\omega_\nu\approx 0.02$.
So, when the likelihood is computed relative to the global minimum
in parameter space allowing $\Omega_{\rm tot}$ to vary, the limit
on the neutrino mass remains unchanged. 
We also find that 
the introduction of massive neutrinos always increases $\chi^2$
relative to the case of massless neutrinos; the presence of massive neutrinos
do not modify the limit on the curvature. 
We finally note that 
the limit on massive neutrinos becomes tightened when 
$\Omega_{\rm tot}\gtrsim 1.03$.

\begin{table}
\begin{tabular}{|c|c|c|c|c |c|c||c||c|c |c|c|}
\hline
$\omega_{\nu}$ & $\omega_b$ & $\omega_m$ &   $h$ & $\tau$ & $n_s$ & $A$
& $\chi^2$ &  $\ell_1$ &  $H_1$ &  $H_2$ & $H_3$ \\
\hline
0.00 & 0.0230 & 0.145 & 0.608 & 0.1222 & 0.973 & 1048.5 & 1427.6 & 220 & 6.92 & 0.448 & 0.454 \\
0.02 & 0.0214 &	0.133&	0.515&	0.0873&	0.910	&	1098.9
&1430.4	&	219	&6.40	&	0.443	&	0.424	\\
0.025& 0.0217 &0.133&		0.504&	0.0878&	0.905& 1119.9 &	1432.4		&
219&	6.31	&	0.440	&	0.423 \\	
0.03
&0.0217&0.126&		0.501&	0.0865	&0.890	&	1173.2&	1434.6&219&	6.16	&	
0.438	&	0.413	\\
\hline
\end{tabular}
\caption{Solutions for $\chi^2_{\rm min}(\omega_{\nu})$ in the
positive curvature universe with $\Omega_{{\rm tot}}=1.02$.}
\label{tab:chi2min_closed}
\end{table}

\begin{table}
\begin{tabular}{|c|c|c|c|c |c|c||c||c|c |c|c|}
\hline
$\omega_{\nu}$ & $\omega_b$ & $\omega_m$ &   $h$ & $\tau$ & $n_s$ & $A$
& $\chi^2$ &  $\ell_1$ &  $H_1$ &  $H_2$ & $H_3$ \\
\hline
0.02&0.0220&0.139	&	0.640&	0.0878&	0.923&		1127.1&
1431.5	&	219		&6.23&		0.442	&	0.435	\\
0.025 &0.0220	&0.134	&	0.627&	0.0871&	0.912&		1146.9&
1433.5	&	219		&6.15	&	0.440	&	0.428	\\
0.03 &0.0220	&0.129	&	0.624&	0.0790&	0.900		&1171.4 &
1435.7	&	219	&	6.06	&	0.439	&	0.420	\\
\hline
\end{tabular}
\caption{Solutions for $\chi^2_{\rm min}(\omega_{\nu})$ in the
negative curvature universe with $\Omega_{{\rm tot}}=0.98$.}
\label{tab:chi2min_open}
\end{table}

It is easy to see how the effect of massive neutrinos 
is modified from the case of the
flat universe. We first note that the partial derivatives
with respect to $\Omega_{\rm tot}$
\begin{eqnarray}
\Delta \ell_1 &=& -360 \frac{\Delta \Omega_{\rm tot}}{\Omega_{\rm tot}}
\label{eq:nonflat_deriv_l1}, \\
\Delta H_1 &=& +4.5\frac{\Delta \Omega_{\rm tot}}{\Omega_{\rm tot}}
\label{eq:nonflat_deriv_H1}.
\end{eqnarray}
The first relation shows the well-known dependence on $\Omega_{\rm tot}$
that the last scattering surface is magnified in the positive curvature
universe. The second
relation arises from the late integrated Sachs-Wolfe effect. For 
$\Omega_{\rm tot}>1$
the reduction of the late integrated Sachs-Wolfe effect decreases
$C_{10}$, and hence increases $H_1$.
$H_2$ and $H_3$ do not depend on $\Omega_{\rm tot}$.
At a first glance one might suspect that
a large response of $\ell_1$ to $\Omega_{\rm tot}$ for $\Omega_{\rm tot}<1$
would cancel the negative change of $\ell_1$ induced by a finite
neutrino mass and relax the limit for
the negative curvature universe. This, however, is not the case.

The position of $\ell_1$ is tightly constrained by the data. So the
change in $\ell_1$ from either the massive neutrino or the departure
 from the flat space curvature is compensated by the change in $h$
that is unconstrained. The negative curvature
makes this shift smaller, and the positive makes it larger as seen
in Figure 2 (c).  
Note that among the 6 cosmological parameters, only $h$ receives
a significant change when a small curvature is introduced. All other
parameters change no more than a few percent from the values for
the flat universe.
The positive curvature increases $H_1$ via 
Eq.~(\ref{eq:nonflat_deriv_H1})
and an extra decrease of $h$ also increases
$H_1$. The increase of $H_1$ makes some
more allowance to the observational lower limit of $H_1$, 
which lowers $\chi^2$ and would in principle weakens the constraint.
However, when we remove the spatial flatness assumption, the global
$\chi^2$ minimum, realised at $\omega_\nu=0$, occurs at a $\chi^2$ smaller
than that for the flat universe.  This offsets the decrease of
$\chi^2(\omega_\nu\ne0)$, and we obtain
the limit on the neutrino mass virtually unchanged from the case
for the flat universe.

Although the limit on the neutrino mass is formally unchanged in
a positive curvature universe, the cost is a significant decrease
of $h$ as seen in Figure 2 (c). 
To realise the 2 $\sigma$ limit, $\omega_\nu\sim 0.021$, we are led to
$H_0\approx 50$ km s$^{-1}$Mpc$^{-1}$, an unacceptably small value.

The argument may go in the opposite way for the negative curvature,
but the limit on the neutrino mass becomes substantially stronger. 
We calculate $\Delta \chi^2$ in a full non-zero curvature parameter space:
for negative curvatures $\Delta \chi^2(\omega_\nu=0)$ is already significant 
relative to the global minimum that is realised in a positive 
curvature universe.

Note that our discussion does not deal with $H_2$ and $H_3$, because
these quantities depend on neither $\Omega_{\rm tot}$ or $h$ directly. The
change of these quantities takes place only through the adjustment of
other parameters, and is small.

In conclusion we find that the constraint on the neutrino mass
we obtained for the flat universe $\omega_{\nu} < 0.021$ is unchanged
even when a non-zero spatial curvature is allowed.

\section{Conclusion} \label{sec:conclusion}
We showed that the subelectronvolt upper limit can be derived
on the neutrino mass from the CMB data alone within the $\Lambda$CDM
model with adiabatic perturbations.
This is contrary to the statements made in
Elgar\o y and Lahav \cite{Elgaroy:2003yh} and
Tegmark et al.\cite{Tegmark:2003ud}, who stressed
that the large-scale galaxy clustering
information is essential to derive the limit on the neutrino mass.
Assuming the flatness of the universe, the constraint we obtained from the one-year data of the WMAP observation alone by maximising the likelihood
    is  $\omega_{\nu} < 
0.021$
or  $\sum m_{\nu} <  2.0$ eV at the 95\% confidence level
(for the degenerate neutrinos, which are close
to the reality if the neutrino mass is close to the limit,
$m_{\nu} <  0.66$ eV).
This is slightly weaker than the limit $<1.7$ eV \cite{Tegmark:2003ud}
derived by the combined use of WMAP and SDSS data, or
similar limits that are obtained by combining more different
types of data \cite{Spergel:2003cb,
Elgaroy:2003yh,Hannestad:2003xv,Allen:2003pt,
Crotty:2004gm,Seljak:2004xh}, 
but our limit is a robust result in the sense that
it does not receive any systematics from biasing, non-linear effects
and others, and solely determined by the CMB data for which systematics
are controlled very well.
Our constraint is unchanged even if we relax the flatness assumption.
The inclusion of the tensor perturbation only tightens the limit.
The assumption we still need is the power-law primordial fluctuation
spectrum.

We argued that it would not be easy to improve the limit beyond
$\sum m_{\nu}\lesssim  1.5$ eV using the CMB data alone, even if
the CMB multipole data are substantially improved. This ``critical
limit'' corresponds to the situation when neutrino becomes 
nonrelativistic
at recombination epoch. That is, we can derive the constraint when
neutrinos become nonrelativistic before the recombination epoch.
The improvement
of the limit on the neutrino mass requires some external inputs,
most characteristically the lower limit on the Hubble constant,
or those that effectively
leads to the constraint on the Hubble constant, such as the Type Ia
supernova Hubble diagram or the large-scale clustering of galaxies.
If $H_0$ would receives a firm lower limit, say
$H_0>65$ km s$^{-1}$Mpc$^{-1}$, the upper limit on the neutrino mass
would be tightened to   $\sum m_{\nu} <  0.8$ eV.

We demonstrated the mechanism as to how these constraints are derived,
using the reduced CMB observables, $\ell_1$, $H_1$, $H_2$ and
$H_3$ introduced by \cite{Hu:2000ti}, and studying their responses to
the neutrino mass density. The key point is that $\ell_1$ and
$H_2$ are constrained to narrow ranges by observation, and the
variation of the cosmological parameters induced by the finite
neutrino density cannot be accommodated in the error budget
of $H_1$ with the increase of the neutrino
mass beyond  $\sum m_{\nu}\sim 2$ eV.

We also showed that the response of the reduced CMB observables,
in particular $\ell_1$ and $H_1$,
to the neutrino mass density is understood by the
modification of the integrated Sachs-Wolfe effect in the presence
of massive neutrinos. In addition, free streaming of massive neutrinos
promotes the decay of gravitational potential that enhances $H_2$ and
$H_3$, whose scales are within free streaming \cite{Dodelson:1995es}.
This leads to the negative correlation between $n_s$ and $m_\nu$,
in contrast to the positive correlation expected from the suppression
of the small scale power due to massive neutrinos.

The most important message from our analysis is that (i) one can
derive the upper limit on the neutrino mass, which is only slightly
weaker than is quoted in the modern literature, using the CMB
(WMAP) data alone: hereby, one can avoid to make use of the
mixed data of different quality or with possible systematic
effects such as biasing and nonlinear effects for galaxies,
and (ii) one may improve the limit by a modest amount even
when the quality of the CMB data is improved, but not much.
For a substantial improvement of the limit one needs a constraint
on the Hubble constant from below.

\appendix
\section{Multidimensional $\chi^2$-minimization} 
\label{sec:minimization}
Our problem is to minimise $f=\chi^2(n_s,\omega_m,\omega_b,\tau,h,A)$ in
6-dimensional parameter space.
Since we want to avoid to calculate the derivative, we adopt
the Brent method \cite{brent} and generalise it to
a multidimensional problem.
For one dimensional problem the Brent method samples 3 points,
$f(x_a),f(x_c),f(x_b)$, and draw a parabola that connects
the three $f$'s to find the value $x_1$ that give the valley of $f$.
Then $f(x_1)$ and the
two neighbouring $f$'s are used to find the next parabola
and its valley at $x_2$. This process is successively applied
until desired convergence.

For multidimensional problem, say, $f(x,y,z)$,
we first minimise $f$ with respect to $z$, by applying the
Brent method in this direction,
with $x$ and $y$ fixed to an arbitrary grid $x_a$ and $y_a$.
We find successively new $z$ grids $z_1(x_a,y_a)$, $z_2(x_a,y_a)$, ...,  
and eventually reach
$f(x_a,y_a,z_{\rm min}(x_a,y_a))$.
We next minimise it with respect to $y$ using
$(y_a, y_b, y_c)$. We carry out the $z$ minimisation for $y_b$ and $y_c$,
i.e., $f(x_a,y_b,z_{\rm min}(x_a,y_b))$ and
$f(x_a,y_c,z_{\rm min}(x_a,y_c))$,
and successively adding a new $y$ grid, $y_1, y_2 ...$, 
while repeating the $z$ minimisation procedure at each step; we 
eventually arrive at 
$f(x_a,y_{\rm min}(x_a),z_{\rm min}(x_a, y_{\rm min}(x_a)))$.
We repeat the same procedure with respect to $x$. 
Starting from 
\begin{eqnarray}
& &f(x_a,y_{\rm min}(x_a),z_{\rm min}(x_a,y_{\rm min}(x_a))), \nonumber 
\\
& &f(x_b,y_{\rm min}(x_b),z_{\rm min}(x_b,y_{\rm min}(x_b))), \nonumber 
\\
& &f(x_c,y_{\rm min}(x_c),z_{\rm min}(x_c,y_{\rm min}(x_c))), \nonumber 
\end{eqnarray}
we finally find
\begin{eqnarray}
f(x_{\rm min},y_{\rm min}(x_{\rm min}),z_{\rm min}(x_{\rm min},
y_{\rm min}(x_{\rm min}))), \nonumber
\end{eqnarray}
which is the desired result.

For our problem of $f=\chi^2(n_s,\omega_m,\omega_b,\tau,h,A)$,
applying the minimisation in the order of 
$A,h,\tau,\omega_b,\omega_m$ and $n_s$,
the final value
would be $\chi^2(n_{s,\rm min}, \omega_{m,\rm min},\omega_{b,\rm 
min},\tau_{\rm min},
h_{\rm min}, A_{\rm min})$, where the omitted arguments are
\begin{eqnarray}
\omega_{m,\rm min} &=& \omega_{m,\rm min}(n_{s,\rm min}), \\
\omega_{b,\rm min} &=& \omega_{b,\rm min}(n_{s,\rm min}, \omega_{m,\rm 
min}), \\
\tau_{\rm min} &=& \tau_{\rm min}(n_{s,\rm min}, \omega_{m,\rm 
min},\omega_{b,\rm min}), \\
h_{\rm min} &=& h_{\rm min}(n_{s,\rm min},
\omega_{m,\rm min},\omega_{b,\rm min},\tau_{\rm min}), \\
A_{\rm min} &=& A_{\rm min}(n_{s,\rm min},
\omega_{m,\rm min},\omega_{b,\rm min},\tau_{\rm min},h_{\rm min}).
\end{eqnarray}

We find that this nested one-dimensional minimizations works well
for the WMAP $\chi^2$ function and the minimum obtained gives $\chi^2$
lower than
those found by the Markov chain Monte Carlo methods given in the
literature. A caution is needed for the
outermost nest, the minimization with respect to $n_s$.
We find two minima for a small $\omega_{\nu}$. So we apply
the minimisation procedure for each case separately.
If more than one mininum is found in the course of 
intermediate minimisation, we must divide the parameter space 
and the  minimisation procedure must be applied separately.
We do not find, however, such cases other than that quoted above.

\section{Comparison of the grid search and MCMC}

\begin{figure}
\includegraphics{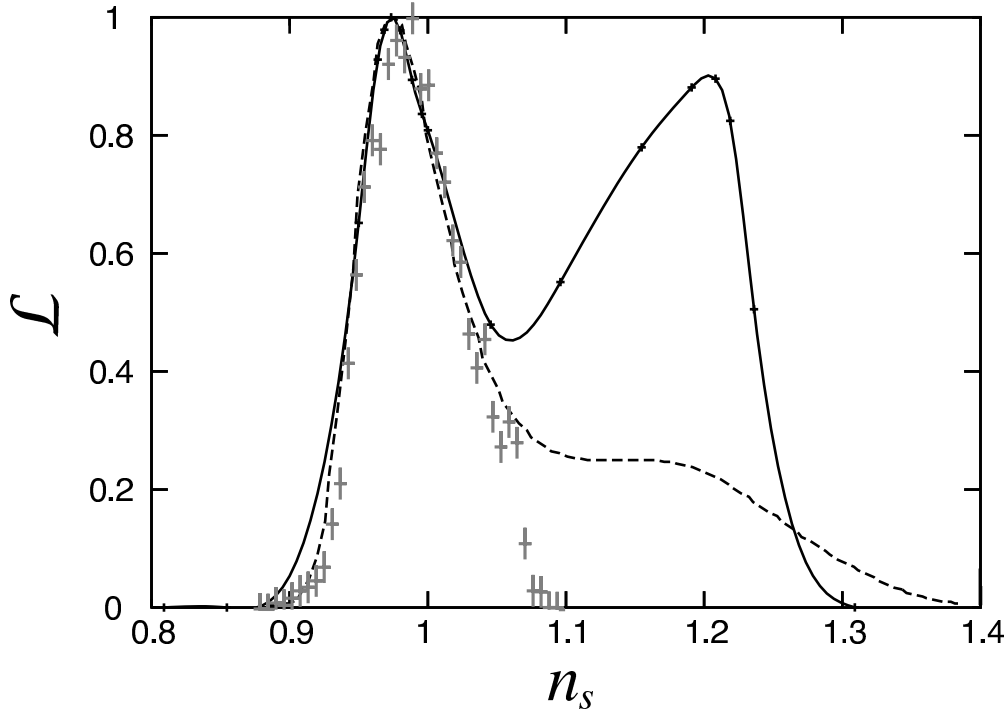}
\caption{Likelihood functions ($\omega_\nu=0$)
for $n_s$ estimated from our $\chi^2$ statistics
(solid line), as compared with those from MCMC given by the WMAP 
group (data points with errors) and Tegmark et al.~(dashed line). 
The maximum is normalised to unity.
}
\label{fig:ns_comparison}
\end{figure}

\begin{figure}
\includegraphics{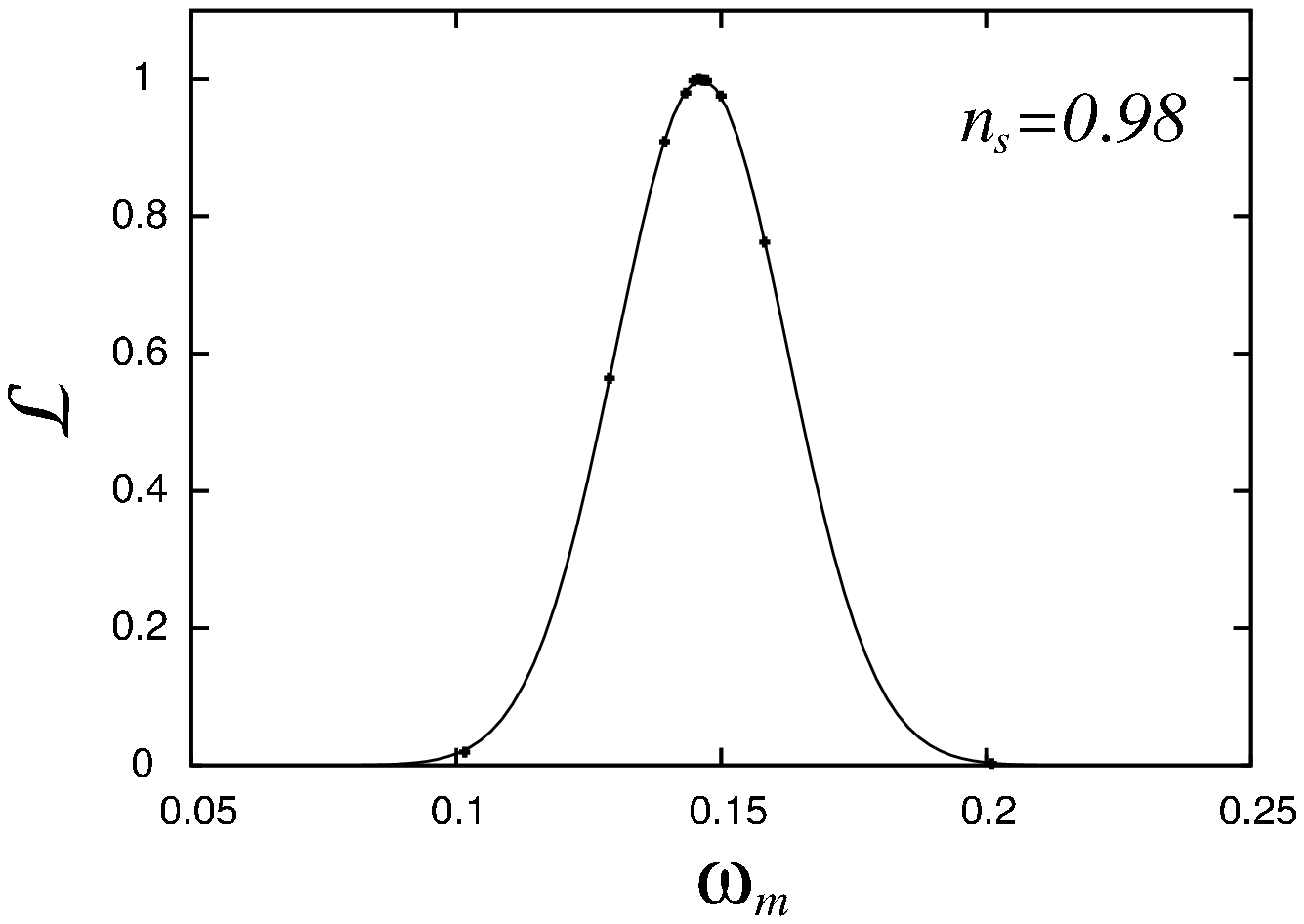}
\caption{An example of the likelihood function for 
$\omega_m$ with $n_s$ fixed to 0.98 and $\omega_\nu$ to zero. 
The 10 data points are fitted with a Gaussian function.
The maximum is normalised to unity.
}
\label{fig:om-like_fix_ns}
\end{figure}

We compare the likelihood function for $\omega_\nu=0$
inferred from the $\chi^2$ function
with those obtained by the MCMC given in the literature. 
In Figure \ref{fig:ns_comparison} we present ${\cal L}=\exp(-\Delta\chi^2/2)$
and $\cal L$ given by Tegmark et al.~\cite{Tegmark:2003ud} 
and Spergel et al.~\cite{Spergel:2003cb} for the variable 
$n_s$.
We see that our likelihood function agrees very well with
Tegmark et al.'s for $n_s<1.05$, but it starts deviating for $n_s>1.05$,
where our likelihood function is much larger, meaning that Tegmark et al.'s
chain does not find a true local minimum near the second peak. 
We emphasise that the relative heights of the two peaks of our $\cal L$
are verified to be close to the `true' likelihoods by marginalising 
the parameters using the multidimensional integral, as mentioned in the text.
The likelihood function of Spergel et al. also agrees with the two
curves. The difference is that they do not get the second peak due to
the prior of $\tau<0.3$.

Figure \ref{fig:om-like_fix_ns} demonstrates an example of the distribution of $\omega_m$,
when $n_s$ is fixed at 0.98. The figure shows a distribution well fitted with
a Gaussian function ($\exp\{ -(\omega_m-a)^2 /2b^2 \}$ with $a=0.146$ and $b=0.0162$). Once one requires the parameters to stay close to one of the
local minima, the distribution is consistent with Gaussian.
This is also true for other 4 parameters. 


\end{document}